\documentclass[a4paper]{article}

\usepackage[english]{babel}
\usepackage[utf8x]{inputenc}
\usepackage[T1]{fontenc}
\usepackage{setspace}

\usepackage[a4paper,top=3cm,bottom=2cm,left=3cm,right=3cm,marginparwidth=1.75cm]{geometry}

\usepackage{lineno}
\usepackage{hyperref}
\hypersetup{colorlinks = true, allcolors = blue}
\usepackage{amsmath}
\usepackage{graphicx}
\usepackage{rotating}
\usepackage[colorinlistoftodos]{todonotes}
\title{FCC-ee: Your Questions Answered \\ {\small Contribution to the European Particle Physics Strategy Update 2018-2020}}
\author{\small (See next page for the list of authors)} 
\date{%5 June 2019
}

\begin{document}
\maketitle

\begin{abstract}
This document answers in simple terms many FAQs about FCC-ee, including comparisons with other colliders. It complements the FCC-ee CDR~\cite{Benedikt:2651299} and the FCC Physics CDR~\cite{Mangano:2651294} by addressing many questions from non-experts and clarifying issues raised during the European Strategy symposium in Granada, with a view to informing discussions in the period between now and the final endorsement by the CERN Council in 2020 of the European Strategy Group recommendations. This document will be regularly updated as more questions\footnote{Send your questions to \href{mailto:patrick.janot@cern.ch,alain.blondel@cern.ch}{patrick.janot@cern.ch} and \href{mailto:alain.blondel@cern.ch,patrick.janot@cern.ch}{alain.blondel@cern.ch}} appear or new information becomes available. 
\end{abstract}

\begin{figure}[ht]
\centering
\includegraphics[width=0.75\textwidth]{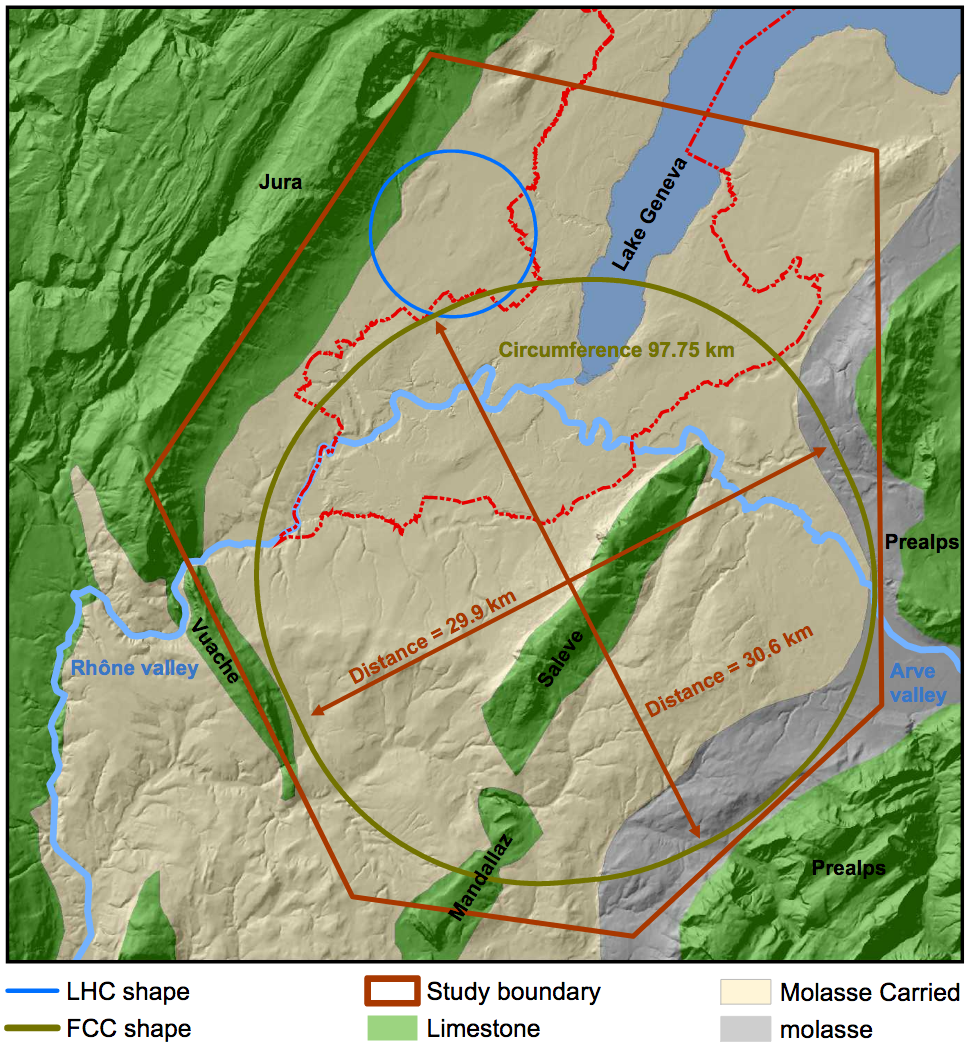}
\caption{\label{fig:PlanMasse} \small Baseline FCC tunnel layout with a perimeter of 97.5\,km, and optimized placement in the Geneva basin, showing the main topographical and geological features.}
\end{figure}

\vfill\eject

\begin{flushleft}
{\large
{\bf This document has been prepared by}
\vspace{0.3cm}

\textsc{A.~Blondel$^{1,2}$,\,
%C.~Grojean$^{2,3}$,\,
P.~Janot$^{2}$ (Editors) 
\vspace{0.3cm} \\
%M.~McCullough$^{4}$\,(editors);\\
%---
%{\bf With contributions from} \\ 
{%\linespread{5}
\begin{spacing}{1.1}
N.~Alipour\,Tehrani$^{2}$,\,
P.~Azzi$^{3}$,\,
P.~Azzurri$^{4}$,\,
N.~Bacchetta$^{3}$,\,
M.~Benedikt$^{2}$,\,
%C.~Bernet$^{7}$,\,
F.~Blekman$^{5}$,\,
M.~Boscolo$^{6}$,\,
M.~Dam$^{7}$,\,
S.~De\,Curtis$^{8}$,\,
D.~d'Enterria$^{2}$,\,
J.~Ellis$^{9}$,\,
%A.~Freitas$^{11}$,\,
G.~Ganis$^{2}$,\,
J.~Gluza$^{10,11}$,\,
%J.~Gu$^{2}$,\,
%B.~Hegner$^{4}$,\, 
%S.~Heinemeyer$^{xx}$,\,
C.~Helsens$^{2}$,\,
S.~Jadach$^{12}$,\,
%J.~Kamenik$^{12}$,\,
%Hamzeh~Khanpour$^{20,21}$,\,
%S.~Khatibi$^{21}$,\,
%M.~Khatiri$^{21}$,\,
M.~Koratzinos$^{13}$,\,
%M.~Mohammadi Najafabadi$^{21}$,\,
M.~Klute$^{13}$,\,
%N.~van~der~Kolk$^{4}$,\, 
%E.~Leogrande$^{4}$,\, 
C.~Leonidopoulos$^{14}$,\,
E.~Locci$^{15}$,\,
M.~Mangano$^{2}$,\, 
S.~Monteil$^{16}$,\, 
K.~Oide$^{2}$,\,
V.\,A.~Okorokov$^{17}$,\,
E.~Perez$^{2}$,\, 
%K.~Peters$^{2}$,\,
%F.~Piccinini$^{16}$,\, 
%M.~Pierini$^{4}$,\,
T.~Riemann$^{10,18}$,\,
%C.~Rogan$^{17}$,\,
%G.~Rolandi$^{4}$,\,
R.~Tenchini$^{4}$,\,
M.~Selvaggi$^{2}$,\,
%F.~Simon$^{22}$,\,
%P.~Skands$^{18}$,\,
%K.~Skovpen$^{19}$,\,
G.~Voutsinas$^{2}$,\, 
J.~Wenninger$^{2}$,\, 
F.~Zimmermann$^{2}$.
\end{spacing}
}}}
\end{flushleft}

\begin{itemize}

\item[$^{1}$] 
  University of Geneva, CH-1205 Geneva, Switzerland
 
% \item[$^{2}$] 
%  DESY, Notkestra{\ss}e 85, D-22607 Hamburg, Germany
 
% \item[$^{3}$] 
%  Humboldt Universit\"at zu Berlin, Newtonstra{\ss}e 15, D-12489 Berlin, Germany
  
\item[$^{2}$] 
  CERN, 1 Esplanade des Particules, CH-1217 Meyrin, Switzerland 

\item[$^{3}$] 
  INFN, Sezione di Padova, Via Marzolo 8, 35131 Padova, Italy

%\item[$^{7}$] 
%  Institut de Physique Nucl\'eaire de Lyon, CNRS/IN2P3, 69622 Villeurbanne, France

\item[$^{4}$] 
  INFN, Sezione di Pisa, Largo Bruno Pontecorvo, 3, 56127 Pisa, Italy

\item[$^{5}$]
  Interuniversity Institute for High Energies, Vrije Universiteit Brussel, Pleinlaan 2,\\ 
  1050 Brussels, Belgium

\item[$^{6}$] 
  INFN, Laboratori Nazionali di Frascati, Via Enrico Fermi 40, 00044 Frascati, Italy

\item[$^{7}$] 
  Niels Bohr Institute, University of Copenhagen, Blegdamsvej 17,\\ 
  2100 Copenhagen, Denmark
  
\item[$^{8}$]
  INFN, Sezione di Firenze, and Department of Physics and Astronomy, University of Florence,\\ Via G. Sansone 1, 50019 Sesto Fiorentino, Italy 

\item[$^{9}$]
  Physics Department, King's College London, Strand, London WC2R 2LS, UK

\item[$^{10}$] 
  Institute of Physics, University of Silesia, 40-007 Katowice, Poland

\item[$^{11}$] 
  Faculty of Science, University of Hradec Králové, Czech Republic
  
%\item[$^{11}$]
%  University of Pittsburgh, Department of Physics and Astronomy, 3941 O'Hara St,\\ 
%  Pittsburgh, PA 15260, USA

%\item[$^{12}$] 
%  Josef Stefan Institute, Theory Department, Jamova cesta 39, 1000 Ljubljana, Slovenia

\item[$^{12}$]
  Institute of Nuclear Physics PAN, ul. Radzikowskiego 152, 31-342 Krak\'ow, Poland

\item[$^{13}$]
  Massachusetts Institute of Technology, 77 Massachusetts Ave,\\ 
  Cambridge, MA 02139, USA

\item[$^{14}$] 
  University of Edinburgh, Department of Physics and Astronomy, Old College,\\ 
  South Bridge, Edinburgh EH8 9YL, UK

\item[$^{15}$] 
  CEA/DRF/IRFU/DPhP, Gif-sur-Yvette \& Université Paris-Saclay, France

\item[$^{16}$]
  Laboratoire de Physique de Clermont, CNRS/IN2P3, 63178 Aubi\`ere Cedex, France%

%\item[$^{16}$] 
%  Universit\`a di Pavia and INFN, Sezione di Pavia, Via A. Bassi, 6, 27100 Pavia, Italy

%\item[$^{13}$]
%  Harvard University, Cambridge, MA 02138, USA

%\item[$^{18}$]
%  Monash University, School of Physics and Astronomy, Clayton VIC 3800, Australia

%\item[$^{20}$]
%University of Science and Technology of Mazandaran, P.O.Box 48518-78195, Behshahr, Iran

%\item[$^{21}$]
%  Institute for Research in Fundamental Sciences (IPM),\\ P.O.Box 19395-5531, Tehran, Iran
  
%\item[$^{22}$]
%Max-Planck-Institut f\"ur Physik (Werner-Heisenberg-Institut) \\
%F\"ohringer Ring 6, 80805 M\"unchen, Germany

%\item[$^{23}$]
%Commissariat~\`{a}~l'Energie~Atomique~et~aux~Energies~Alternatives (CEA) Saclay, 91191 Gif-sur-Yvette Cedex, France

%\item[$^{24}$]
%Universit\'{e}~Paris-Saclay, 91190 Saint-Aubin, France
  
\item[$^{17}$]
  National Research Nuclear University MEPhI (Moscow Engineering Physics Institute),\\
  Kashira Highway, 31, Moskva, Russia, 115409  
  
\item[$^{18}$]
  Deutsches Elektronen–Synchrotron, DESY, 15738 Zeuthen, Germany

\end{itemize}

\vfill\eject

\tableofcontents

\vfill\eject

\parskip=0.15cm
\section*{Editors' note}

{\it With the discovery of the Higgs boson, the matrix of particles and interactions described by the "Standard Model" (SM) is complete. 
This consistent and predictive theory has so far been successful at  describing all phenomena accessible to collider experiments. 
Why should particle physics continue?

Particle physics has indeed arrived at an important moment in its history. The discovery of the Higgs boson with a mass of 125\,GeV opens a new era, clearing the decks for a new phase of exploration of physics beyond the Standard Model. For the first time since the Fermi theory, there is no clear guide what form this may take. Several fundamental experimental facts, however, remain unexplained, such as the abundance of matter over antimatter, the evidence for dark matter, and the non-zero neutrino masses. Many theoretical issues also require physics beyond the present Standard Model (BSM). Particle physics must continue its investigations, in the broadest possible way, with a mix of radical improvements in sensitivity, precision, and energy range.

\noindent The physics programme of an ambitious post-LHC accelerator complex must address the following goals. 
\begin{itemize}
    \item  Map the properties of the Higgs and electroweak gauge bosons, pinning down their interactions
with an accuracy order(s) of magnitude better than today, and acquiring sensitivity to, e.g. the processes
that, during the time span from $10^{-12}$ and $10^{-10}$ s after the Big Bang, led to the formation of today's Higgs
vacuum field.
\item Improve by at least an order of magnitude the discovery reach for new particles at the highest masses. 
\item Improve by orders of magnitude the sensitivity to rare and elusive phenomena at low energies, including the possible discovery of particles with very small couplings. In particular, the search for dark matter should seek to reveal, or conclusively exclude, dark matter candidates belonging to broad classes of models, such as weakly-interacting massive particles with a relic density fixed during the thermal history of the universe.
\item Probe energy scales beyond the direct kinematic reach, via an extensive campaign of precision measurements,
%, searches for forbidden or rare decays and symmetry violations, 
sensitive to tiny deviations from the Standard Model behaviour. This campaign requires high event statistics, exquisitely precise experimental conditions, and improved theoretical calculations, as well as the maximal amount of synergies within the programme.           
\end{itemize}

The analysis of community positions presented at the recent open European Strategy meeting as well as the discussions, indicated that {\it (i)} the next machine ought to be an $e^+e^-$ collider; {\it (ii)} Europe should proceed with a flagship collider programme at CERN; and {\it (iii)} a vigorous R\&D programme must continue to pave the way towards the highest possible centre-of-mass energy with high luminosities. 

The FCC design study has shown that, taking into account today's physics landscape, an integrated Future Circular Collider programme at CERN consisting of a luminosity-frontier $e^+e^-$ collider (FCC-ee) followed by an energy-frontier hadron collider (FCC-hh, with an FCC-eh option), both very strongly motivated in their own right, promises the most far-reaching particle physics programme that foreseeable technology can deliver. The FCC also offers, as is the case for the LHC, an outstanding and diversified programme of interest for the nuclear and heavy-ion physics community. The unique opportunities provided by the FCC injector complex and the continuation of the proton accelerators at CERN are  further benefits of the FCC integrated programme. Given the overall scale, duration and cost of each of the future collider projects, it is indeed crucial to keep alive smaller projects and open the participation to the widest possible physics community.

The FCC-ee is a rather young, rapidly evolving, project, which has raised a number of questions prior, during, and after the European Strategy symposium in Granada (May 2019). We hope that compiling these questions, together with their answers, in a single reference document will clarify issues and inform the scientific and strategic discussions between now and the final endorsement by the CERN Council in 2020 of the European Strategy Group recommendations. A more complete overview of the FCC physics  programme can be found in the FCC CDRs~\cite{FCC-CDR}, in particular the description of the FCC integrated project~\cite{Benedikt:2653673}, the FCC physics CDR~\cite{Mangano:2651294}, the FCC-ee CDR~\cite{Benedikt:2651299}  and the FCC-hh CDR~\cite{Benedikt:2651300}.
}

\section{What is FCC-ee?}
\begin{figure}[ht]
\centering
\includegraphics[width=0.50\textwidth]{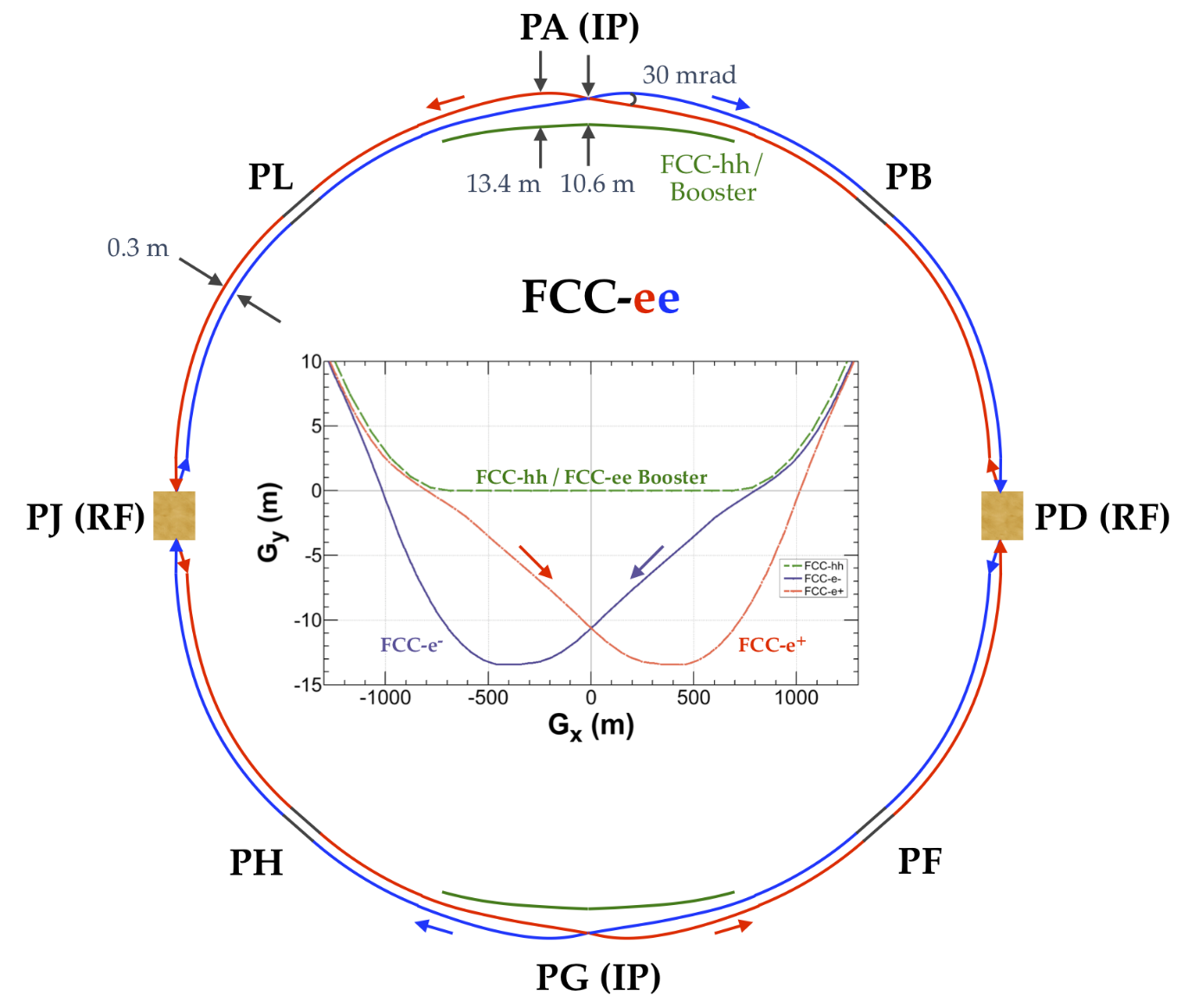}
\includegraphics[width=0.47\textwidth]{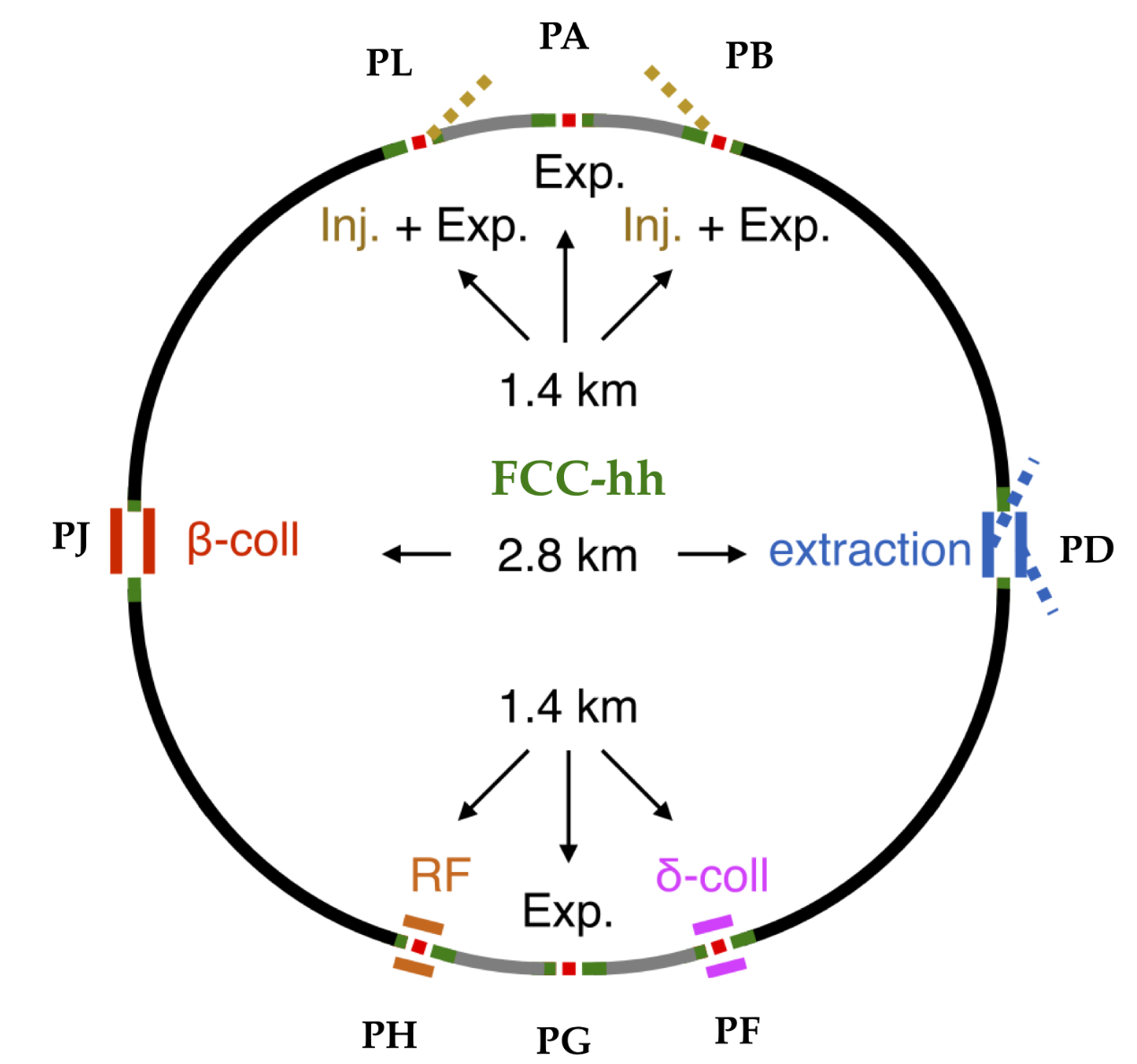}
\caption{\label{fig:Schematic} \small Schematics of the implementation of the FCC-ee collider (left) and the FCC-hh collider (right) in the common infrastructure. The FCC-ee booster footprint coincides with that of the FCC-hh.~The asymmetric $\rm e^\pm$ beam lines around the FCC-ee interaction regions are designed to minimize synchrotron radiation in the detectors.}
\end{figure}

The FCC-ee~\cite{Benedikt:2651299} is the first stage of the integrated Future Circular Colliders (FCC) programme~\cite{Benedikt:2653673}, to be based on a novel research infrastructure hosted in a $\sim 100$\,km tunnel in the neighbourhood of CERN, as illustrated in Fig.~\ref{fig:PlanMasse}.~Most of the FCC-ee infrastructure can be directly re-used for a subsequent energy-frontier ($\sim 100$\,TeV) hadron collider (FCC-hh)~\cite{Benedikt:2651300} -- and possibly offer opportunities for the realization of energy-frontier (3 to 14\,TeV) muon colliders -- providing the world-wide particle-physics community with multiple interaction points (IPs) -- two IPs for FCC-ee and four IPs for FCC-hh in the current baseline design shown in Fig.~\ref{fig:Schematic} -- in a highly synergistic and cost-effective manner during the 21{st} century.  

\vskip -.3cm

\vskip -.3cm
\begin{figure}[h]
\centering
\includegraphics[width=0.70\textwidth]{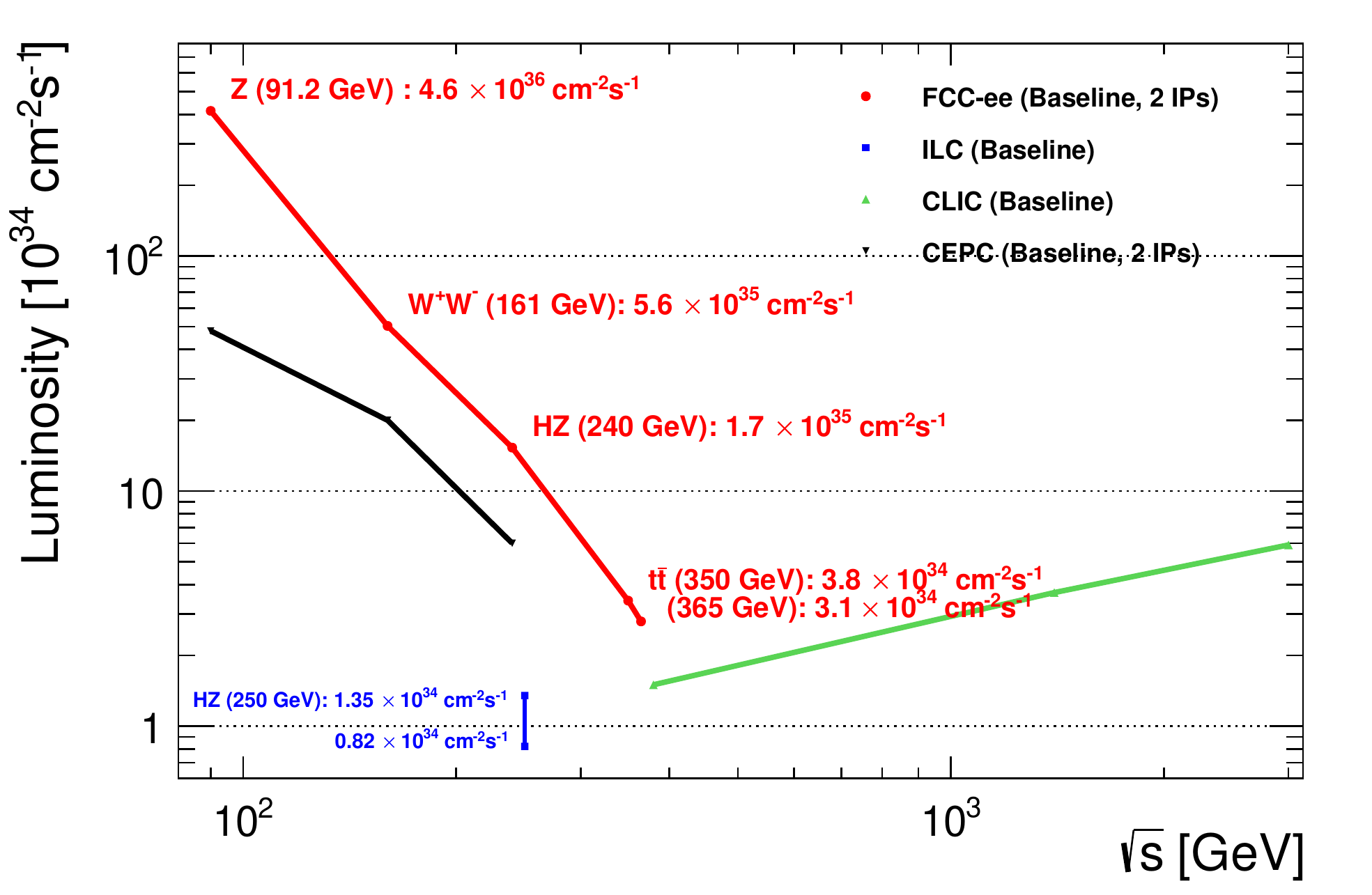}
\caption{\label{fig:OperationModel} \small The FCC-ee baseline design luminosity~\cite{Benedikt:2651299} summed over 2 IPs as a function of the centre-of-mass energy $\sqrt{s}$, compared to the baseline luminosities{\color{blue}$^2$} of other ${\rm e^+e^-}$ collider proposals (ILC~\cite{Adolphsen:2013kya,Evans:2017rvt}, CLIC~\cite{Robson:2018enq}, and CEPC~\cite{CEPCStudyGroup:2018rmc}.)}
\end{figure}

{\bf The FCC-ee is the high-energy ${\bf e^+e^-}$ collider project with the highest luminosity proposed to date in its baseline design}\footnote{It is technically possible to double the baseline luminosities of all the ${\rm e^+ e^-}$ colliders
considered here, either by doubling the rate of the positron source at a linear collider~\cite{Bambade:2019fyw}, or by serving four IPs (Section~\ref{sec:TwoInteractionPoints}) instead of two at a circular collider. The two values for the ILC baseline luminosity correspond to the TDR~\cite{Adolphsen:2013kya} value ($0.82 \times 10^{34}\,{\rm cm}^{-2}{\rm s}^{-1}$), and the new baseline~\cite{Evans:2017rvt} value ($1.35 \times 10^{34}\,{\rm cm}^{-2}{\rm s}^{-1}$), obtained by reducing further the beam sizes with respect to the TDR (Section~\ref{sec:BeamSizes}).}, as shown in Fig.~\ref{fig:OperationModel}.
The design is inspired by the progress made for B factories with top-up injection, strong focusing, and crab-waist optics, allowing luminosities $10^5$ times larger than at LEP~\cite{Assmann:2002th} to be achieved. The FCC-ee will be implemented in stages as an {\bf electroweak, flavour, and Higgs factory} {to study with unprecedented precision the Higgs boson, the Z and W bosons, the top quark, and other particles of the Standard Model}, by spanning the energy range from the Z pole and the WW threshold through the maximum Higgs production rate, up to the ${\rm t\bar t}$ threshold and beyond, in a 15-year experimental programme summarized in Table~\ref{tab:OperationModel}.

\begin{table}[ht!]
\begin{center}
\caption{\small The baseline FCC-ee operation model, showing the centre-of-mass energies, instantaneous luminosities for each IP, integrated luminosity per year summed over the 2 IPs corresponding to 185 days of physics per year and 75\% efficiency. As a conservative measure, the yearly integrated luminosity is further reduced by 10\% in this table, and in all physics projections. The total luminosity is set by the physics goals, which in turn set the run time at each energy. The luminosity is assumed to be half the design value for commissioning new hardware during the first two years at the Z pole and in the first year at the ${\rm t\bar t}$ threshold.\vspace{0.4cm} \label{tab:OperationModel}}
\begin{tabular}{|l|c|c|c|c|c|c|}
\hline 
Working point & Z, {\footnotesize years 1-2} & Z, {\footnotesize later} & WW & HZ & \multicolumn{2}{|c|}{${\rm t\bar t}$} \\ \hline
$\sqrt{s}$ {\footnotesize (GeV)} & \multicolumn{2}{|c|}{88, 91, 94} & 157, 163 & 240 & {\small 340-350} & 365 \\ \hline
{\small Lumi/IP {\footnotesize ($10^{34}\,{\rm cm}^{-2}{\rm s}^{-1}$)}} & 115 & 230 & 28 & 8.5 & 0.95 & 1.55 \\ \hline
{\small Lumi/year {\footnotesize (${\rm ab}^{-1}$, 2 IP)}} & 24 & 48 & 6 & 1.7 & 0.2 & 0.34 \\ \hline
Physics Goal {\footnotesize (${\rm ab}^{-1}$)} & \multicolumn{2}{|c|}{150} & 10 & 5 & 0.2 & 1.5 \\ \hline
Run time {\footnotesize (year)} & 2 & 2 & 2 & 3 & 1 & 4 \\ \hline
 & \multicolumn{2}{|c|}{} & & $10^6$ \,HZ & \multicolumn{2}{|c|}{$10^6 {\rm t\bar t}$} \\
Number of events &  \multicolumn{2}{|c|}{$5\times 10^{12}$ Z} & $10^8$ WW & + & \multicolumn{2}{|c|}{$+200$k HZ} \\
 & \multicolumn{2}{|c|}{} & & 25k WW $\to$ H & \multicolumn{2}{|c|}{$+50$k\,${\rm WW}\to {\rm H}$} \\ \hline
\end{tabular} 
\end{center}
\end{table}

{
\section{Can I do Higgs physics in the first year of FCC-ee?}
In the baseline scenario shown in Table~\ref{tab:OperationModel} above, the first six years of FCC-ee would be devoted to Z and W physics, whereas other lepton colliders propose an operation model starting with Higgs physics at 240, 250, or 380\,GeV. While the  proposed  staging  plan for FCC-ee is  the  most natural, efficient, economic, and logical operation model from the point of view of  the  machine  installation  and  evolution, there is no technical obstacle to run first at $\sqrt{s}= 240$\,GeV: {\bf flexibility is one of the many strong points of the FCC integrated programme.}

If the run at 240\,GeV were to be the first stage in the FCC-ee operation model, it would last four years instead of three, to allow for two years of machine commissioning with half the design luminosity, and the later Z pole run would then last three years instead of four. During the initial four years, regular short runs at the Z pole will be scheduled to align and calibrate the detectors (Section~\ref{sec:ExperimentalUncertainties}), each offering several ``{Giga-Z's}\,'' for first-stage precision electroweak measurements. Alternatively, the FCC-ee schedule is flexible enough to start with two years at 240\,GeV -- which would already provide Higgs coupling measurements comparable to other proposed low-energy Higgs factories (Section~\ref{sec:LowEnergyHiggsFactory}) and render model-independent all HL-LHC measurements -- and then go back to the baseline programme (Z, WW, HZ, etc.) or to any variant to be discussed and agreed upon in due time by the physicists working on the FCC-ee experiments. 

}

\section{How can the FCC-ee Machine Parameters reach such High Luminosities?}

The question was posed back in 2011, when the first parameter lists were proposed in Ref.~\cite{Blondel:2011fua}, with luminosities up to $10^4$ times larger than at LEP. Subsequently there have been in-depth studies, incorporating the latest advances in B-factory design, and the conclusion is that {\bf FCC-ee can be built, with even better performance than originally thought.} 

\subsection{What is the basis for the FCC-ee machine parameters?}

The FCC-ee design builds up on 50 years of experience with circular ${\rm e^+e^-}$ colliders, and exploits the historical knowledge from

\begin{itemize}
\item 
LEP: A large ring, capable of high energy, using Nb/Cu RF cavities, with strong synchrotron radiation, energy calibration by spin resonance, and a high beam-beam parameter (in excess of $0.1$);
\item 
SLC: Strong ${\rm e^+e^-}$ sources with damping rings;
\item 
VEPP-4: Precise energy calibration using beam polarization (Section~\ref{sec:Polarization});
\item 
KEKB and PEP-II B factories and BEPC-II: Two separate rings for electrons and positrons to allow for the largest number of bunches, continuous (top-up) injection, mitigation of electron cloud effects, highest stored electron current (PEP-II), non-zero crossing angles (KEKB and BEPC-II), and high beam-beam parameters (larger than $0.1$);
\item 
DA$\Phi$NE: Crab-waist optics to optimize the collision area with a crossing angle, and the highest stored current for positrons;
\item 
Super B factories (Super KEK-B): Strong focusing with small $\beta^*$, $L^*$, and large Piwinski angle;
\end{itemize}
and combines this experience with a few recent, novel ingredients (twin dipole and quadrupole magnets, novel vacuum chamber design to control the synchrotron radiation load (50 MW/beam), pure Nb RF cavities to reach the highest energies, asymmetric $\rm e^\pm$ beam lines around the interaction regions to minimize synchrotron radiation in the detectors,  etc.), to reach extremely high luminosities at high energies. 

The parameters are now supported by a detailed design study, during which complete and independently cross-checked simulations of many features were performed, 
such as off-momentum dynamic aperture; beam-beam instabilities; flip-flop effect; crab-waist optimization; working-point optimization; beamstrahlung and beam lifetime; bootstrapping for first full injection; injector cycle and minimum sustainable lifetime; etc. {\bf As a result, the performance of the FCC-ee machine presented in the Conceptual Design Report~\cite{Benedikt:2651299} has improved beyond what the original assumptions would offer, and the parameters are much more robust: this collider can be built with the proposed luminosities.} 

The independent CEPC study has led to luminosities that confirm the anticipated FCC-ee performance. The slightly smaller luminosities provided by the CEPC design (typically a factor of two) are mostly due to a smaller design beam synchrotron power (30 MW/beam instead of 50 MW/beam); a larger defocusing effect at the IPs (mostly at the Z pole) due to the larger detector magnetic field (3\,T instead of 2\,T in one of the two detectors); and different working-point optimization (to limit the beamstrahlung effect on the beam energy spread). 

\subsection{How do circular and linear \texorpdfstring{${\bf e^+e^-}$}{ee} colliders compare in this respect?}

As mentioned above, the technology and accelerator physics of circular ${\rm e^+e^-}$ colliders are based on a considerable amount of experience and are supported by a broad community. Several linear colliders have been proposed since the early 1990's, and have been studied in extensive simulations and paper studies, as well as through accelerator R{\&}D. These have led to the publication of a TDR, while the FCC-ee and CEPC proposals are currently at the level of CDR. The operational experience with linear colliders, however, is limited and a number of issues that are critical for luminosity are still in R{\&}D phase. Some known issues are listed below.

\subsubsection{\it Historical record}

All recent high-energy lepton circular colliders, such as LEP1~\cite{Assmann:2002th,Myers:226776}, LEP2~\cite{Assmann:2002th}, PEP-II \cite{Seeman:1078536} and KEKB \cite{Akai:2001pf}, achieved their design peak luminosities within (less than) one to five years after start-up. They also all ultimately exceeded their design luminosities by factors of 2 to 4.
{\bf For this reason, the design luminosities of future circular colliders may be considered as conservative estimates.} For example, during the KEKB design phase, there were warnings that the KEKB design luminosity would never be achieved~\cite{bib:kimura2010}, but it was exceeded by more than a factor of two by the end of the programme.  

On the other hand, the only linear collider so far, SLC, reached half of its design peak luminosity only after 10 years of operation \cite{slc-c,Assmann:2000xv}. Its beam commissioning proved a considerable challenge. Linear colliders are single-pass systems with slightly different beam parameters on every pulse (pulse-to-pulse ``jitter''~\cite{Adolphsen:1995nx}), and long-term drifts.  The luminosity performance relies on achieving small collision spot sizes that are to be accomplished by intricate final-focus systems and instrumentation, with  continuous optics tuning and feedbacks~\cite{Emma:1997cp}. In the most successful last year of SLC running (1998), the average  vertical beam size at the collision point was still two-and-a-half times larger than the beam size expected from the incoming beam parameters and the final-focus optics~\cite{Zimmermann:1998fx}.

\subsubsection{\it Beam sizes}
\label{sec:BeamSizes}

The two linear-collider final-focus test facilities FFTB (1994-1997) and ATF2 (since 2009) achieved, after extended tuning periods, unprecedentedly small world-record spot sizes; yet the minimal vertical spot sizes were still about two to eight times larger than expected for the respective optics and measured input emittances~\cite{fftb,atf2,marcin}. The ATF2 spot size was decreased below 100 nm after reducing the bunch charge by five to ten and increasing $\beta_{x}^{\ast}$ ten times compared with the design~\cite{atf2}. The lower bunch charge not only reduces wake-field effects, but also decreases the transverse emittance and the energy spread, which are dominated by intra-beam scattering in the damping ring~\cite{atf-dr}. The much-larger-than-design value of $\beta_{x}^{\ast}$ suppresses the important geometric and chromo-geometric aberrations affecting the vertical spot size. With these relaxed running conditions, the expected ATF2 rms vertical beam size for the modified emittance and optics would be less than 30\,nm, i.e., noticeably smaller than the 37\,nm design value. If ATF2 were a collider, the above changes in bunch intensity and $\beta_{x}^{\ast}$ would amount to a luminosity reduction by a factor of  $\sim 100$.

\begin{figure}[htbp]
\centering
\includegraphics[width=0.47\textwidth]{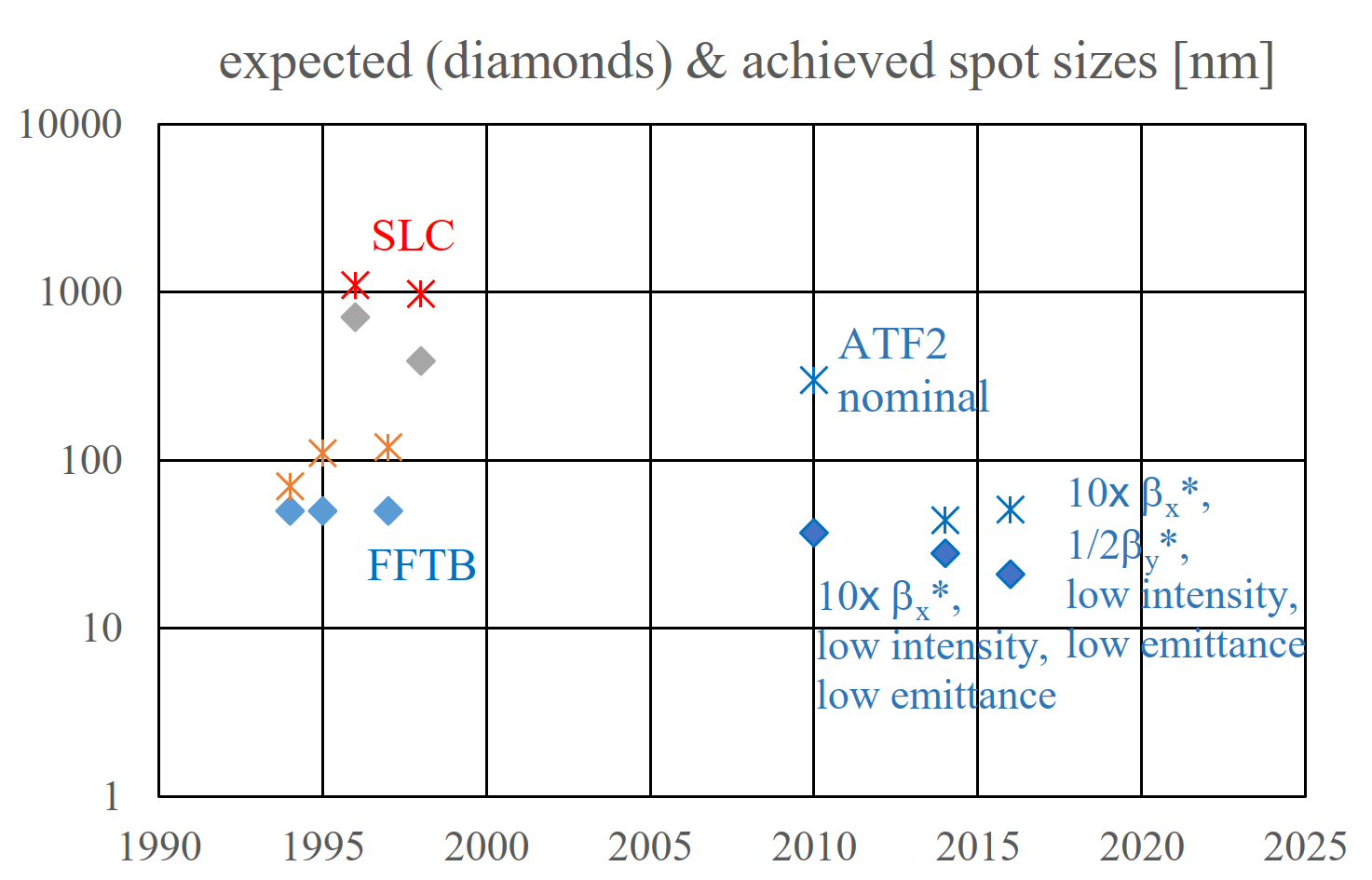}
\includegraphics[width=0.51\textwidth]{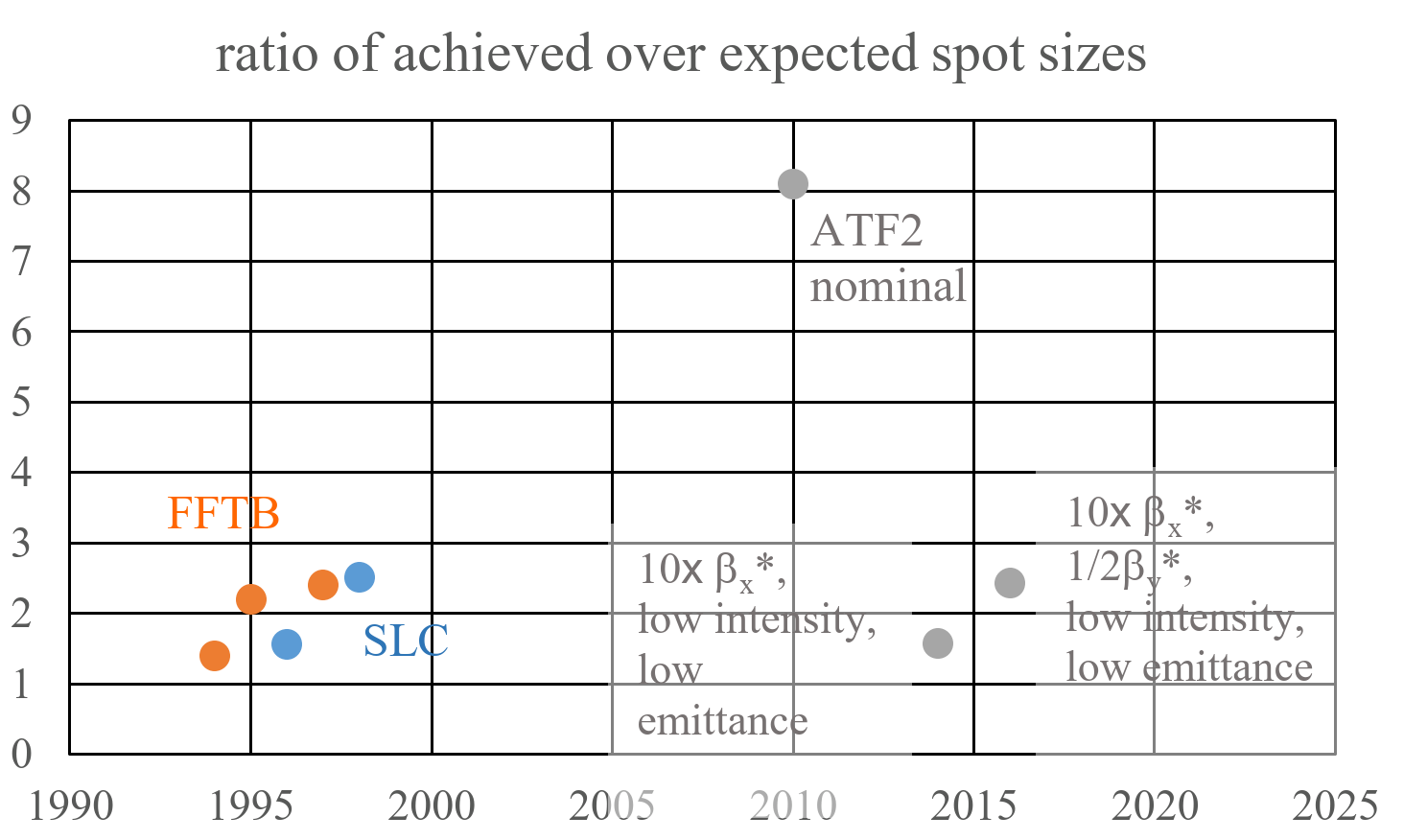}
\caption{\label{fig:spotsize} \small Expected and typically achieved, or average, vertical rms 
spot sizes at the FFTB \cite{fftb}, SLC \cite{Zimmermann:1998fx,Zimmermann:1996fx} and ATF2 \cite{marcin,atf2,steinar} (left) and the ratio of the achieved and expected spot sizes (right).}
\end{figure}

The reasons for, and the various contributions to, the larger-than-expected spot sizes at SLC, FFTB and ATF2 are not yet entirely understood~\cite{Zimmermann:1998fx,fftb,marcin}.  Examples of large differences between  expected and achieved beam sizes at FFTB, SLC, and ATF2 are shown in Fig.~\ref{fig:spotsize}. In view of this experience, it would seem prudent to assume that the average collision-point spot size of a future linear collider might be significantly larger than the design value, by a factor of two or more. 

\subsubsection{\it Positron source}
All present and future circular colliders require positron production rates that are significantly lower than the production rates routinely achieved at the SLC and expected from the positron injector of SuperKEKB. 

On other other hand, linear colliders require positron sources with production rates that exceed the present world record (established at SLC) by factors of 20 to 40. The ILC baseline polarized positron source relies on a novel scheme of high-energy photon conversion, which needs the full-energy electron beam. The density of energy deposition in the target is very high and requires a rotating target scheme, with serious issues of radiation and cooling. This point is critical for the combined possibility of luminosity upgrades and polarized positron beams and has not been demonstrated yet.  
%The sealing of the rotating target is challenging, and its feasibility has not yet been  
% NB the latest design used cooling pads, so no seal is needed. 
Post-TDR R{\&}D is on-going on the positron target scheme and interesting progress is made~\cite{ilc-eplus}, but the full validation will have to wait for the availability of the full energy and intensity electron beam, i.e., the construction of at least half of the entire collider. 

\subsubsection{\it Beam emittance }

Finally, while low-emittance electron beams have been stored and maintained in storage rings for decades, and are presently experiencing a renaissance with the advent of ultimate diffraction-limited storage-ring light sources, a linear collider requires extracting a low-emittance beam from a storage ring, which is not a standard mode of operation. At both SLC and KEK ATF, the extraction of bunches from the damping ring appeared to result in a noticeable emittance increase compared with the parameters measured in the ring prior to extraction~\cite{slc-ex,Alabau:2008zz,atf-ex2}. This emittance blow-up during 
extraction was attributed to nonlinear fields experienced in the magnets passed far off-centre, to imperfect kicker pulses, to stray fields 
in the extraction septum, and to wake fields.

\subsection{Summary}
The above observations certainly comfort us that the circular collider technology is a reliable and relatively low-risk choice for future lepton colliders.
%, when it comes to luminosity promises. Yet, in quite a few places, 
Nevertheless, additional optimization and detailed design of several aspects of the FCC-ee are planned for the next five years, with the aim of reducing the cost and the risks, of further checking and improving the performance, and of finalizing a number of choices. Those are discussed in Section~\ref{sec:ReadyToGo}.

\section{How will the FCC-ee Detectors deal with Beam Backgrounds?}
\label{sec:BeamBackgrounds}

This question has arisen in view of the high design luminosity (100,000 times larger than at LEP when running at the Z pole). It should be recalled, however, that this luminosity is spread over a large number of bunches (e.g., 16,640 bunches when running at the Z pole, compared to 4 bunches at LEP), the intensity of each being similar to that of the LEP bunches. In addition, the asymmetric design of the interaction point seen in Fig.~\ref{fig:Schematic} is such that the synchrotron radiation at the highest energies is not very different from that at LEP2, so that related backgrounds in the detectors were found to be negligible.  

Detailed simulations of synchrotron radiation, incoherent ${\rm e^+e^-}$ pair production and hadron photo-production, combined with a GEANT-4 simulation of the asymmetric interaction region and of the two detector concepts studied in the CDR, show that beam backgrounds are under control at FCC-ee and, for all practical purposes, negligible. For example, at each bunch crossing at FCC-ee, the occupancy in a typical vertex detector is found to be smaller than $10^{-5}$ at the Z pole, and a few $10^{-4}$ at 365\,GeV. Even with a very slow readout electronics integrating over 1\,$\mu$s (corresponding to 50 bunch crossings at the Z pole), the maximum occupancy  observed  would  remain well below $10^{-3}$.  

All things considered, {\bf it was demonstrated during the FCC-ee design study that detectors satisfying the requirements are feasible.} The experience built up over decades of linear collider detector studies has been instrumental in this process. Future developments in the coming decade can only improve further on this already favourable situation. 

\section{How good is FCC-ee as a Higgs Factory?}
\label{sec:LowEnergyHiggsFactory}

The FCC-ee is an excellent low-energy ${\rm e^+e^-}$ Higgs factory indeed, complementing the LHC and synergistic with FCC-hh. The FCC-ee is also much more than an excellent Higgs factory. This second aspect is developed in Section~\ref{sec:MuchMore}. 

First, it is interesting to note that FCC-ee delivers to its two interaction points, at 240\,GeV (the HZ rate maximum), a baseline luminosity that is larger by a factor of 12.6 
(resp.\,11.3) than the ILC (resp.~CLIC) baseline luminosity at 250 (resp.~380)\,GeV. On the other hand, the HZ production cross section is larger at ILC by a factor of 1.19, thanks to the choices of longitudinal beam polarization ($\pm 80\%$ for the electron beam, $\pm 30\%$ for the positron beam, with 45\% of the luminosity in each of the $\pm \mp$ polarization configurations and the rest in the $\pm\pm$ configurations). On average, there is no such cross-section boost for CLIC, as the CLIC positron beam is assumed to be unpolarized\footnote{The CLIC team has concluded that the physics returns from positron polarization are not worth the trouble, risk, and money necessary to get significant positron polarization in collisions.}. In addition, CLIC suffers from a smaller HZ cross section (by a factor of 1.8) due to the larger centre-of-mass energy. All in all, FCC-ee is therefore expected to collect 10 (20) times more Higgs bosons every year at 240\,GeV than ILC (CLIC) at 250 (380)\,GeV with the HZ process. The FCC-ee also foresees a five-year run at the ${\rm t\bar t}$ threshold and above (an energy scan between 340 and 350\,GeV for a year, then 365\,GeV for four years) with a luminosity that is twice that of CLIC at 380\,GeV, thanks to the two FCC-ee interaction points. 

The {\it Higgs@FutureColliders} working group, set up in view of the Symposium for the European Strategy Update in Granada, came to the results~\cite{deBlas:2019rxi} displayed in Table~\ref{tab:kappaEFT} for the projected Higgs coupling precisions at the different facilities, {\it (i)} in the $\kappa$ framework (allowing for H coupling strength multipliers); and {\it (ii) } in a global EFT fit to Higgs, diboson, and electroweak precision measurements. 
\begin{table}[ht]
\begin{center}
\caption{\label{tab:kappaEFT} \small Precision on the Higgs boson couplings, as determined by the {\it Higgs@FutureColliders} working group~\cite{deBlas:2019rxi} in the $\kappa$ framework (left) and in a global EFT fit (right), for the four low-energy Higgs factories (ILC, CLIC, CEPC, and FCC-ee), fixing the Higgs self-coupling $g_{\rm HHH}$ to its SM value. For $g_{\rm HHH}$ fit, the first number assumes that all other couplings are fixed to their SM values. All numbers are in \% and indicate 68\% C.L. sensitivities. Also indicated in the $\kappa$ fit are the precision on the total decay width and the 95\% C.L. sensitivity on the "invisible" and "exotic" branching fractions, the latter accounting for final states that cannot be tagged as SM decays. All numbers include current projected parametric uncertainties, and are combined with the projected HL-LHC precision~\cite{Cepeda:2650162} given in the first column. In the $\kappa$ fit, the HL-LHC result is obtained by fixing the total Higgs boson width and the ${\rm H \to c\bar c}$ branching fraction to their Standard Model values, and by assuming no BSM decays. A specific ${\rm H\to Z} \gamma$ or ${\rm e^+e^- \to H}\gamma$ analysis has so far been attempted only by the CEPC team. The HZ$\gamma$ coupling is otherwise obtained solely from HL-LHC projections.\vspace{0.4cm}}
\begin{tabular}{|l|c|c|c|c|c|}
%\multicolumn{6}{c}{} \\ 
\hline Collider & {\small HL-LHC} & {\small ILC$_{250}$} & {\small CLIC$_{380}$} & {\small CEPC$_{240}$} & FCC-ee$_{240\to 365}$ \\ \hline
Lumi (${\rm ab}^{-1}$) &  {\small 3} & {\small 2} & {\small 1} & {\small 5.6} &  5 + 0.2 + 1.5 \\ \hline
Years & & {\small  11.5}\,$^5$ & {\small 8} & {\small 7} & 3 + 1 + 4 \\ \hline
$g_{\rm HZZ}$ (\%) &  {\small 1.5 / 3.6} & {\small 0.29 / 0.47} & {\small 0.44 / 0.66}  & {\small 0.18 / 0.52} & {\bf 0.17 / 0.26}  \\ %\hline
$g_{\rm HWW}$ (\%) &  {\small 1.7 / 3.2} & {\small 1.1 / 0.48 } & {\small 0.75 / 0.65} & {\small 0.95 / 0.51} & {\bf 0.41 / 0.27}   \\ %\hline
$g_{\rm Hbb}$ (\%) &  {\small 3.7 / 5.1} & {\small 1.2 / 0.83} & {\small 1.2 / 1.0} & {\small 0.92 / 0.67} & {\bf 0.64 / 0.56} \\ %\hline
$g_{\rm Hcc}$ (\%) &  {\small SM / SM} & {\small 2.0 / 1.8} & {\small 4.1 / 4.0} & {\small 2.0 / 1.9} & {\bf 1.3 / 1.3} \\ %\hline
$g_{\rm Hgg}$ (\%) &  {\small 2.5 / 2.2} & {\small 1.4 / 1.1} & {\small 1.5 / 1.3} & {\small 1.1 / 0.79} & {\bf 0.89 / 0.82}  \\ %\hline
$g_{\rm H\tau\tau}$ (\%) &  {\small 1.9 / 3.5} & {\small 1.1 / 0.85} & {\small 1.4 / 1.3} & {\small 1.0 / 0.70} & {\bf 0.66 / 0.57} \\ %\hline
$g_{\rm H\mu\mu}$ (\%) &  {\small 4.3 / 5.5} & {\small 4.2 / 4.1} & {\small 4.4 / 4.3} & {\small 3.9 / 3.8} & {\bf 3.9 / 3.8} \\ %\hline
$g_{\rm H\gamma\gamma}$ (\%) &  {\small 1.8 / 3.7} & {\small 1.3 / 1.3} & {\small 1.5 / 1.4} & {\small 1.2 / 1.2} & {\bf 1.2 / 1.2} \\ %\hline
$g_{\rm HZ\gamma}$ (\%) &  {\small 11. / 11.} & {\small 11. / 10.} & {\small 11. / 9.8} & {\small 6.3 / 6.3} & {\bf 10. / 9.4} \\ %\hline
$g_{\rm Htt}$ (\%) &  {\small 3.4 / 2.9} & {\small 2.7 / 2.6} & {\small 2.7 / 2.7} & {\small 2.6 / 2.6} & {\bf 2.6 / 2.6} \\ \hline
$g_{\rm HHH}$ (\%) &  {\small 50. / 52.} & {\small 28. / 49.} & {\small 45. / 50.} & {\small 17. / 49.} & {\bf 19. / 34.} \\ \hline
$\Gamma_{\rm H}$ (\%) &  {\small SM} & {\small 2.4 } & {\small 2.6} & {\small 1.9} & {\bf 1.2}  \\ \hline
BR$_{\rm inv}$ (\%) &  {\small 1.9} & ${\small 0.26}$ & {$\small 0.63$} & ${\small 0.27}$ & ${\bf 0.19}$ \\ 
BR$_{\rm EXO}$ (\%) &  {\small SM (0.0)} & ${\small 1.8}$ & {$\small 2.7$} & ${\small 1.1}$ & ${\bf 1.0}$ \\ \hline
\end{tabular} 
\end{center}
\end{table}

{\bf The numbers in this Table speak for themselves: the FCC-ee is not "just another" Higgs factory. It is the best of the low-energy Higgs factory candidates on the table.}

Each of the low-energy factories compared in Table~\ref{tab:kappaEFT}, if realized, is expected to be continued with an energy-frontier collider\footnote{In the case of ILC, the energy upgrades to 500\,GeV and to 1\,TeV are absent from the project currently discussed with the Japanese government.}, with ${\rm e^+e^-}$ collisions for CLIC and ILC, and with proton-proton collisions in the CEPC and FCC tunnels. The prices of all these energy-frontier upgrades are in the 20-billion range (Section~\ref{sec:DeadEnd}), which is too expensive to be realized in the short term, while the costs of the low-energy Higgs factories (Section~\ref{sec:cost}) are deemed to be affordable as the next big collider for particle physics. The durations of the Higgs programmes are also comparable, typically eight years\footnote{In the case of ILC, the duration of 11.5 years~\cite{Bambade:2019fyw} is based on the hypotheses {\it (i)} that a doubling of the luminosity will occur after five years; and {\it (ii)} that a year will provide $1.6\times 10^7$ seconds for physics. The latter is significantly larger than the $1.2\times 10^7$ and $1.08\times 10^7$ seconds assumed for CLIC and FCC-ee, by 33\% and 50\%, respectively, and also larger than any of the ${\rm e^+e^-}$ colliders that have been operated so far~\cite{PhysRevD.54.1,PhysRevD.86.010001}. The last, most successful, and by far the longest run of the SLC (which lasted 10 months in 1997-1998, almost without any technical interruption) delivered 350,000 Z to SLD with an average peak luminosity of 150-200 Z/hour~\cite{Phinney:2000av}, barely reaching $10^7$ effective seconds/year. The circular collider record is held by the BELLE detector operating at KEKB in 2006~\cite{Brodzicka:2012jm,Kimura:2012xua}, with 190\,${\rm fb}^{-1}$ collected at a peak luminosity of 16\,${\rm nb}^{-1}/{\rm s}$, corresponding to almost $1.2\times 10^7$\,seconds/year. With the same year definition as for FCC-ee, the ILC$_{250}$ run duration would be 16.5 years, increasing to 18.5 years in the absence of a luminosity upgrade.}.  The comparable price tag and duration are the reasons why only low-energy Higgs factories are compared in Table~\ref{tab:kappaEFT}. The energy-frontier upgrades are compared in Section~\ref{sec:EnergyUpgrades}. 
    
\section{How Many Interaction Points at FCC-ee?}
\label{sec:TwoInteractionPoints}

The configuration of the FCC-ee with two interaction points was chosen by the FCC international steering board in 2015. 

This configuration offers a stimulating competition between the corresponding two collaborations, in both experimental approach and physics exploitation. Most importantly, it provides the possibility to independently confirm observations that would hint at new physics,
which is essential in view of the history of unconfirmed discoveries made by a single particle physics experiment. This is a major advantage of circular colliders. Besides, {\bf two interaction points deliver twice as much luminosity as a single interaction point. This is another substantial advantage when precision is the name of the game.}

In this context, the obvious next question is {\bf "Why not four interaction points?"}, as was the case for LEP and as was assumed in the original TLEP paper~\cite{Gomez-Ceballos:2013zzn}. At the time of their decision, the FCC international steering board had wisely concluded that a configuration with four interaction points would be more difficult to validate in the time-limited FCC conceptual design study, and that a bird in the hand was worth two in the bush. After the publication of the CDR, this question has come back, and the CERN management has requested a complementary study with four interaction points. Beside the substantial advantage of serving twice as many users for 15 years, preliminary studies show that the luminosity would be almost doubled with respect to two IPs. Multi-turn simulations are being run to check the beam stability in this configuration, and to evaluate the needs for design modifications.

With the additional flexibility offered by this luminosity increase, an improved strategy could be envisioned: {optimistically}, two years at the Z pole and one year at the WW threshold would suffice, {in principle,} to get almost the same integrated luminosity as after six years with two IPs. The saved years and the four detectors could then be optimally used to accumulate up to about 12\,${\rm ab}^{-1}$ at 240\,GeV and up to 5.5\,${\rm ab}^{-1}$ at 350 and 365\,GeV. {Before the combination with HL-LHC, all FCC-ee Higgs coupling precisions would be improved by a factor of up to 1.7 in this configuration~\cite{cdr-higgs-studies} compared to two IPs (under the assumption that theory uncertainties match the experimental precision, as discussed in Section~\ref{sec:TheoryErrors}).} %are displayed in Table~\ref{tab:FourIP}. 
The larger data sample would in particular yield a model-independent measurement of the Higgs self-coupling with a precision of $\pm 21\%$ (and of $\pm 12\%$ if only $g_{\rm HHH}$ is allowed to vary)~\cite{Blondel:2018aan}. 

{\bf With four interaction points, the first $\bf 5\sigma$ demonstration of the existence, i.e. the discovery, of the Higgs self-coupling is therefore within reach with 15 years of FCC-ee measurements.} {The increased flexibility offered by the four-detector configuration at FCC-ee will offer multiple choices to the future FCC-ee experiment committee, which may give higher priority to other measurements. Another fascinating possibility, for example, would be to spend a couple of years at $\sqrt{s} = 125$\,GeV, with a view to constraining the electron Yukawa coupling~\cite{dEnterria:2017dac}.}

%\begin{table}[h]
%\begin{center}
%\caption{\label{tab:FourIP} \small Precisions on the Higgs boson couplings and on the Higgs boson mass and width at the FCC-ee with four IPs, compared to the baseline projections with two IPs~\cite{cdr-higgs-studies}. No combination with HL-LHC (except for $g_{\rm HHH}$) and no parametric uncertainties are included here.\vspace{0.4cm}}
%%\multicolumn{6}{c}{} \\ 
%\hline Collider & Two IPs & Four IPs \\ \hline
%Lumi (${\rm ab}^{-1}$) &  5 + 0.2 + 1.5 & 12 + 0.2 + 5.3 \\ \hline
%Years & 3 + 1 + 4 & 3.5 + 0.5 + 7.5 \\ \hline
%$g_{\rm HZZ}$ (\%) &  0.17  & 0.10\\ %\hline
%$g_{\rm HWW}$ (\%) &  0.43  & 0.24 \\ %\hline
%$g_{\rm Hbb}$ (\%) &  0.61  & 0.36\\ %\hline
%$g_{\rm Hcc}$ (\%) &  1.21  & 0.73\\ %\hline
%$g_{\rm Hgg}$ (\%) &  1.01  & 0.60 \\ %\hline
%$g_{\rm H\tau\tau}$ (\%) &  0.74 & 0.43\\ %\hline
%$g_{\rm H\mu\mu}$ (\%) &  9.0 & 5.5\\ %\hline
%$g_{\rm H\gamma\gamma}$ (\%) & 3.9 & 2.2\\ \hline
%$g_{\rm HHH}$ (\%) & 34 & 21 \\ \hline
%$\Gamma_{\rm H}$ (\%) & 1.3 & 0.77  \\ \hline
%BR$_{\rm inv}$ (\%) & 0.19 & 0.13\\ 
%BR$_{\rm EXO}$ (\%) & 1.0 & 0.65 \\ \hline
%\end{tabular} 
%\end{center}
%\end{table}

\section{Do we need an \texorpdfstring{${\bf e^+e^-}$}{ee} Energy of at least \texorpdfstring{500\,GeV}{500GeV} to Study the Higgs Boson Thoroughly?}
\label{sec:EnergyUpgrades}

The document submitted to the CERN Council by the European Strategy for Particle Physics (ESSP) Group in 2013~\cite{deliberationDocument} explained that a lepton collider with {\it "energies of 500\,GeV or higher could explore the Higgs properties further, for example the {\rm [Yukawa]} coupling to the top quark, the  {\rm [trilinear]} self-coupling  and  the  total  width."}. Variations on this qualitative argument have been used to argue that an ILC upgrade to 500\,GeV would allow the measurement of the Higgs potential and would increase the potential for new particle searches~\cite{LCBReport,ICFAStatement}. As a consequence, the strategic question was raised again whether the FCC-ee design study ought to consider a 500\,GeV energy upgrade. In the context of this European Strategy Update and in view of the the Granada symposium, the ESSP 2013 statement was revisited quantitatively (Table~\ref{tab:kappaEFT}), and it was found~\cite{Blondel:2018aan}, that
\begin{itemize}
\item The FCC-ee can measure the total width of the Higgs boson with a precision of 1.2\% -- the best precision on the market -- with runs at $\sqrt{s} = 240$, $350$, and $365$\,GeV, and without the need of an energy upgrade to 500\,GeV;

\item The top Yukawa coupling will have been determined at HL-LHC at the $\pm$3.4\% level, albeit with some model dependence, without the need of 500\,GeV ${\rm e^+e^-}$ collisions; and that the combination of this HL-LHC result with the FCC-ee absolute Higgs coupling and width measurements removes the model dependence, without the need of an energy upgrade to 500\,GeV; 

\item The FCC-ee provides a $3\sigma$ sensitivity to the Higgs self-coupling, from the precise measurement of the single-Higgs production cross section as a function of $\sqrt{s}$, and that with four experiments instead of two (Section~\ref{sec:TwoInteractionPoints}), might well achieve the first model-independent $5\sigma$ demonstration of the existence of the Higgs self-coupling, without the need of an energy upgrade to 500\,GeV; 

\item A precise measurement of the Higgs self-coupling at the few per-cent precision level can realistically only be provided by the combination of FCC-ee and FCC-hh, which is beyond the reach of lepton colliders with centre-of-mass energies up to at most 3\,TeV. 

\end{itemize}

{\bf In summary, 500\,GeV is not an essential energy in the context of Higgs boson studies for the lepton colliders under consideration}, especially for FCC-ee. An upgrade of FCC-ee to 500\,GeV would, in addition, have significant impact on the timing and cost of the overall FCC programme, while its physics output would be largely superseded by FCC-hh. The projected Higgs coupling precisions expected with a combination of the low-energy Higgs factories and their proposed energy upgrades are displayed in Table~\ref{tab:EnergyUpgrades}. 
\begin{table}[ht]
\begin{center}
\caption{\label{tab:EnergyUpgrades} \small Precision on the Higgs boson couplings, as determined by the {\it Higgs@FutureColliders} working group~\cite{deBlas:2019rxi} in the $\kappa$ framework (left) and in a global EFT fit (right), for the combination of each low-energy Higgs factory (ILC$_{250}$, CLIC$_{380}$, and FCC-ee) and their proposed upgrades: ILC$_{500\,{\rm GeV}}$, CLIC$_{1.4+3\,{\rm TeV}}$, and FCC-hh+eh. All numbers are in \% and indicate 68\% C.L. sensitivities, and are combined with the projected HL-LHC precision. Also indicated are the precision on the total decay width, and the 95\% C.L. sensitivity on the "invisible" and "exotic" branching fractions. {The rightmost column indicates the time needed (in years) for FCC-hh to reach a precision similar to that offered by the full programme of the proposed linear collider upgrades.} \vspace{0.4cm}}
\begin{tabular}{|l|c|c|cc|}
%\multicolumn{6}{c}{} \\ 
\hline Collider & {\small ILC$_{250+500}$} & {\small CLIC} & FCC & Years\\ \hline
$g_{\rm HZZ}$ (\%) & {\small 0.23 / 0.22}  & {\small 0.39 / 0.20} & {\bf 0.17 / 0.13} & $< 1$ \\ %\hline
$g_{\rm HWW}$ (\%) & {\small 0.29 / 0.23} & {\small 0.38 / 0.18} & {\bf 0.20 / 0.13} & $< 1$  \\ %\hline
$g_{\rm Hbb}$ (\%) & {\small 0.57 / 0.52} & {\small 0.53 / 0.38} & {\bf 0.48 / 0.44} & $< 1$ \\ %\hline
$g_{\rm Hcc}$ (\%) & {\small 1.2 / 1.2} & {\small 1.4 / 1.4} & {\bf 0.97 / 0.95} & $< 1$ \\ %\hline
$g_{\rm Hgg}$ (\%) & {\small 0.84 / 0.79} & {\small 0.86 / 0.75} & {\bf 0.53 / 0.49} & $< 1$ \\ %\hline
$g_{\rm H\tau\tau}$ (\%) & {\small 0.64 / 0.60} & {\small 0.82 / 0.73} & {\bf 0.49 / 0.45} & $< 1$ \\ %\hline
$g_{\rm H\mu\mu}$ (\%) & {\small 3.9 / 3.9} & {\small 3.5 / 3.4} & {\bf 0.44 / 0.42} & $< 1$\\ %\hline
$g_{\rm H\gamma\gamma}$ (\%) & {\small 1.2 / 1.1} & {\small 1.1 / 1.1} & {\bf 0.36 / 0.34} & $<1$ \\ %\hline
$g_{\rm HZ\gamma}$ (\%) & {\small 11. / 6.7} & {\small 5.7 / 3.7} & {\bf 0.70 / 0.70} & $<1$\\ %\hline
$g_{\rm Htt}$ (\%) & {\small 2.4 / 2.4} & {\small 1.9 / 2.0} & {\bf 0.95 / 1.6} & $<1$ \\ \hline
$g_{\rm HHH}$ (\%) & {\small 27./27.} & {\small 11./n.a.} & {\bf 5./6.} & $1$--$7$ \\ \hline
$\Gamma_{\rm H}$ (\%) & {\small 1.4} & {\small 1.6} & {\bf 0.91} & $<1$\\ \hline
BR$_{\rm inv}$ (\%) & {$\small 0.22$} & ${\small 0.61}$ & ${\bf 0.024}$ & $<1$ \\ 
BR$_{\rm EXO}$ (\%) & {$\small 1.4$} & ${\small 2.4}$ & ${\bf 1.0}$ & $<1$ \\ \hline
\end{tabular} 
\end{center}
\end{table}

This Table demonstrates that, while ${\rm e^+e^-}$ energy upgrades indeed bring improvements to some Higgs boson coupling measurements (though not very significantly with respect to the FCC-ee programme), proton-proton collisions are qualitatively and quantitatively more effective to study the Higgs boson thoroughly. {\bf The FCC integrated plan is indeed the only programme leading to precision consistently smaller than 1\% for all couplings to gauge bosons and to fermions and for the Higgs boson total width, and to a precision of few per cent for the Higgs self-coupling.} 

{An important question is how long FCC-hh would require to attain these precisions~\cite{Beate}. As indicated in Table~\ref{tab:kappaEFT}, the FCC-ee precision for most of the couplings of Table~\ref{tab:EnergyUpgrades} is already similar to that of the full programme of the linear collider upgrades, and the FCC-hh would need less than a year to catch up, wherever needed. Regarding the Higgs self-coupling, FCC-hh would also require less than a year to reach the precision projected for ILC$_{500}$ (though about seven years to reach that projected for CLIC$_{3000}$). Therefore, FCC could match the linear collider precisions within a similar time duration, not $> 50$ years as might have been misunderstood from Ref.~\cite{Beate}.}

\section{Why are the FCC-ee Beams not Polarized Longitudinally?}
\label{sec:Polarization}

\subsection{A choice: Longitudinal or Transverse Polarization?}
\label{sec:choice}
Among the physics requirements for the design of FCC-ee, beam polarization -- transverse and longitudinal -- has been carefully considered. At circular colliders, transverse beam polarization builds up spontaneously by the Sokolov-Ternov effect, and enables a very precise calibration of the beam energy by resonant depolarization, at the $\pm 100 $ keV level.  {\bf Such an accurate beam energy calibration is unique to circular lepton (electron and muon) storage rings}. It allows in turn the determination of the Z boson mass and width and of the W mass with the same level of precision. The precision of these measurements is a cornerstone of the electroweak physics programme of FCC-ee, and a long-lasting contribution to the field. In a situation where the Standard Model is "complete", this program is a powerful tool, in particular to search for SM deviations pointing to the existence of further electroweakly-coupled particles.

The interest and possibilities for longitudinal polarization at FCC-ee have also been investigated in great detail. Conceptual schemes can be devised~\cite{Gomez-Ceballos:2013zzn,Koop:2015jpa} for spin rotators, with which  longitudinal polarization could be obtained for both beams symmetrically at some or all centre-of-mass energies. The separate rings for positrons and electrons offer great flexibility for spin manipulations, which -- combined with resonant depolarization -- allow the necessary  calibration of the colliding particle polarization~\cite{Blondel:1987wr}. For particles such as the Z and W bosons and the top quark (which are produced and which decay via parity-violating weak interactions), however, {\bf longitudinal polarization brings no information that could not be obtained otherwise}, e.g., from the study of the final-state kinematics with the large event samples expected at the FCC-ee. A comprehensive discussion can be found in Ref.~\cite{Blondel-berlin} for what concerns Z-pole measurements, and in Refs.~\cite{Janot:2015yza,Janot:2015mqv} for what concerns the top quark coupling measurements. Further arguments are given below for the Z pole in Section~\ref{sec:LongitudinalZ}, and for Higgs coupling measurements in Section~\ref{sec:LongitudinalH}. 

The combination of top-up injection of unpolarized beams and the significant time needed to build up polarization lead to a loss of luminosity and serious constraints on operation: a Z peak run with longitudinal polarization would lead to a fifty-fold loss in luminosity (i.e., $10^{11}$ instead of $5 \times 10^{12}$ Z) for an effective polarization of the beams of 30\% ~\cite{Blondel-berlin}. 
Given the unique and important opportunities offered by the high statistics of the TeraZ physics program (Section~\ref{sec:trillions}); given that the precision of the beam energy calibration could be affected; and given  that the corresponding polarization observables can be obtained otherwise; longitudinal polarization was not considered worth the effort during the conceptual design study. Consequently, detailed study of implementing longitudinal polarization equipment has been considered lower priority and deferred to a later discussion, should a clear physics case emerge.

\subsection{Longitudinal GigaZ vs Transverse TeraZ}
\label{sec:LongitudinalZ}

The following discussion is a response to a question raised at the European Strategy symposium~\cite{Gudi}. The acknowledged need to complete Higgs coupling measurements by Z pole measurements has triggered a renewed interest for the  "GigaZ" program at linear colliders~\cite{Irles:2019xny}, which had somehow been left aside in the recent proposals. This opportunity, which was included in, e.g., the TESLA project~\cite{AguilarSaavedra:2001rg}, is interesting for linear colliders, because a longitudinally polarized electron beam is readily available from the source, with no loss of luminosity. 

Historically, a certain amount of complementarity is indeed visible between the Z pole runs at SLC (with 500,000 Z) and LEP (with 20 million Z), as illustrated in Table~\ref{tab:Gudi}, extracted from the Review of Particle Physics listings~\cite{PDGLive2018}. 
More details can be found in Ref.~\cite{ ALEPH:2005ab}.  It can be seen that SLC benefited greatly from longitudinal electron polarization for the measurement of the effective weak mixing angle, $\sin^2\theta_{\rm eff}^{\rm lept}$; it is not clear, however, how well SLC would have done without the input from LEP for the Z lineshape parameters.    
The SLC also reaped benefits from the smaller beam pipe, which in turn allowed much better flavour tagging in the SLD detector. This latter point is not a concern at FCC-ee, because the clean conditions and the strong focusing allow for a beam-pipe size similar to or even smaller than at linear colliders. We thus concentrate on the potential benefits of longitudinally polarized beams in the following. 

\begin{table}[ht]
\begin{center}
\caption{\label{tab:Gudi} \small Comparisons between the published SLC and LEP precision for a number of representative measurements at the Z pole~\cite{PDGLive2018}. Also indicated are the critical parameters allowing these precisions to be reached. Most of the SLC measurements use the Z mass value provided by LEP, which was enabled by LEP transverse polarization. In the last year of operation, a Z-peak scan was performed to calibrate the SLC spectrometers to the LEP measurement of the Z mass, leading to a total $\sqrt{s}$ uncertainty of 29\,MeV~\cite{Rowson:2001cd}. \vspace{0.4cm}}
\begin{tabular}{|l|c|c|c|}
\hline Measurement & LEP Precision & SLC Precision & Critical parameter \\ \hline
$m_{\rm Z}$ (MeV) &  2.1 & 120  & Transverse polarization at LEP \\ %\hline
$\Gamma_{\rm Z}$ (MeV) &  2.3 & 400 & Transverse polarization at LEP \\ %\hline
$\Gamma_{\rm e^+e^-}$ (MeV) &  0.12 & 1.5 & Statistics at LEP \\ %\hline
$\Gamma_{\rm invisible}$ (MeV) &  1.5 & -- & Statistics at LEP \\ %\hline
$\Gamma_{\rm hadrons}$ (MeV) &  2.0 & -- & Statistics at LEP \\ %\hline
$R_\ell = \Gamma_{\rm hadrons}/\Gamma_{\ell^+\ell^-}$ &  0.025 & 3.4 & Statistics at LEP \\ %\hline
$R_{\rm c} = \Gamma_{\rm c\bar c}/\Gamma_{\rm hadrons}$ &  0.052 & 0.031 & Smaller SLC beam pipe and ...  \\ %\hline
$R_{\rm b} = \Gamma_{\rm b\bar b}/\Gamma_{\rm hadrons}$ &  0.0007 & 0.0012 &  ... better SLD vertex detector \\ \hline
& &  & Longitudinal polarization at SLC ... \\
$\sin^2\theta_{\rm eff}^{\rm lept}$ & 0.00021 & 0.00026 & .. and statistics at LEP  \\ 
& &  & (with asymmetries: $A_{\rm FB}^{\ell\ell}$, $A_{\tau}^{\rm pol}$, ...)  \\ \hline 
\end{tabular} 
\end{center}
\end{table}

The recent work presented in Ref.~\cite{Irles:2019xny} stresses the interest of longitudinal beam polarization in view of stringent lepton universality measurements at a GigaZ. This important aspect is indeed not sufficiently stressed in the FCC-ee CDR~\cite{Benedikt:2651299} and ESPP contribution~\cite{Benedikt:2653669}, and is therefore explicitly addressed here. Lepton universality measurements have been recently discussed in the context of the so-called heavy flavour anomalies~\cite{Graverini:2018riw}, as well as in the search for indirect effects of right-handed neutrinos heavier than the Z~\cite{Antusch:2015mia}. Three different tests of lepton universality can be performed at the Z factory run of the FCC-ee. 
\begin{enumerate}
\item Charged current universality tests are best performed with the $1.7 \times 10^{11}$ $\tau^+\tau^-$ pairs produced during the FCC-ee TeraZ run. Due to the very low level of confusion between $\tau$ pairs and hadronic Z decays at these energies, the Z pole is a particularly interesting place to study $\tau$ decays. As a matter of fact, the leptonic branching ratio measurements at LEP and the associated universality tests are still unsurpassed, in spite of 100 times higher statistics accumulated at B factories~\cite{lepton-universality-HFLAV-2017}. The present tests of $\rm e/\mu/\tau$ universality in tau decays stand at the per-mil level~\cite{lepton-universality-HFLAV-2017}. With the TeraZ run, a precision at the $10^{-5}$ level can be contemplated for these tests. Further tests can be performed in heavy-flavour decays in the $\sim 10^{12}$ Z decays to $\rm b\bar{b}$ and $\rm c\bar{c}$. Here, longitudinal beam polarization does not help: a polarized GigaZ, with three order of magnitude fewer events, is significantly more limited. 

\item Neutral current universality can be tested first from the comparison of the partial widths of the Z into each of the three lepton pairs. The partial width into a lepton pair amounts to 
\begin{equation*}
\Gamma_{\ell\bar{\ell}} = \frac{G_F m_Z^3}{6\pi\sqrt{2}} \left( a_\ell^2 +v_\ell^2 \right).
\end{equation*}
Given the smallness of the lepton vector couplings $v_\ell$, the measurement of the leptonic partial widths is basically a test of universality of the axial-vector couplings $a_\ell$. Because lepton species are well separated from each other, and because such issues as angular acceptance and large angle radiation cancel in the ratios, this test is only limited by the size of the lepton pair samples. It can therefore be expected that this test be performed with a precision better than $10^{-5}$ at the TeraZ; here again the GigaZ will directly suffer from 3000 times less statistics. An absolute determination of the electron partial width can be directly constrained from the measurement of the total Z peak cross section and the knowledge of the Z width,  which is best measured at FCC-ee with the aforementioned outstanding beam energy calibration.  

\item Finally, the ratio of vector-to-axial-vector couplings can be accessed through measurements of initial- and final-state polarization asymmetries, as well as forward-backward asymmetries. For these asymmetry measurements, initial-state beam polarization brings substantial improvements in statistical power, but these measurements can be also performed without, as quantitatively assessed below.  
\end{enumerate} 

The quantities of interest are the left-right coupling asymmetries
\begin{equation*}
{\cal A}_{\rm f} = 
\frac{ g_{\rm L,f}^2- g_{\rm R, f}^2} { g_{\rm L, f}^2 + g_{\rm R, f}^2}
= \frac{2v_{\rm f} a_{\rm f}}{a_{\rm f}^2 +v_{\rm f}^2 }.
\end{equation*}
For leptons, $ {\cal A}_\ell \simeq 2 \left( 1-4\sin^2\theta_{\rm W}^{\rm eff} \right) \simeq 0.15$. For longitudinally polarized $\rm e^+$ and $\rm e^-$  beams, the $\rm e^+e^-$ system acquires a longitudinal polarization ${\cal P} =(P_{\rm e^-} - P_{\rm e^+})/(1-P_{\rm e^-} P_{\rm e^+})$. The spins are measured along the incident particles direction. In the case of, e.g., unpolarized positrons, ${\cal P} = P_{\rm e^-}$. 
With incoming longitudinally-polarized beams, the inclusive beam polarization asymmetry $A_{\rm LR}$ and the forward-backward beam polarization asymmetry $A_{\rm FB, pol}^{\rm f}$ (a definition of these quantities can be found in Refs.~\cite{Blondel:1987wr,ALEPH:2005ab}) can be measured directly:
\begin{equation*}
A_{\rm LR}  =  {\cal P}{\cal A}_{\rm e}\, , {\rm \ \ and \ \ }
A_{\rm FB, pol}^{\rm f}  =  \frac{3}{4} {\cal P}{\cal A}_{\rm f}.
\end{equation*}
Without longitudinally-polarized beams, the same left-right coupling asymmetries can be accessed by using either forward-backward unpolarized asymmetries $A_{\rm FB}^{\rm f}$, or the final-state $\tau$ polarization $\langle P_{\tau} \rangle$ and its forward-backward polarization asymmetry $A_{\rm FB, pol}^{\tau}$: 
\begin{equation*}
A_{\rm FB}^{\rm f}  =  \frac{3}{4} {\cal A}_{\rm e} {\cal A}_{\rm f}\, , \ \
 \langle P_{\tau} \rangle  =  -{\cal A}_{\tau}\, , {\rm \ \ and \ \ }
A_{\rm FB, pol}^{\tau}  =  \frac{3}{4} {\cal A}_{\rm e}\, .
\end{equation*}

The main advantages offered by longitudinal beam polarization are that {\it (i)} $A_{\rm LR}$ can be measured with any visible Z decay, and has thus impressive statistical power; and {\it (ii)} if  ${\cal P} > {\cal A}_{\rm e}$, the sensitivity to the final-state coupling is enhanced by initial-state polarization. The main potential drawback is that the individual measurements with beam longitudinal polarization are limited by the knowledge of the beam polarization. In Ref.~\cite{Irles:2019xny}, a superb beam polarization knowledge of $\Delta{\cal P}/{\cal P} = 5 \times 10^{-4}$ is assumed. For universality tests, it is preferable to measure the ratios, for which this systematic uncertainty cancels. The evaluated precisions\footnote{We find uncertainties that are almost three times larger for ${\cal A}_{\mu}$ and  ${\cal A}_{\tau}$ at the GigaZ than in the version of Ref.~\cite{Irles:2019xny} available at the time of writing, because the factor $\tfrac{3}{4} {\cal P} = 0.6  $ worsens the uncertainties rather than improve them. One cannot measure an asymmetry with a precision of $10^{-4}$ with $3\times 10^7$ muon pairs.} on the left-right coupling asymmetries from the ILC GigaZ collecting  $0.7 \times 10^9$ hadronic Z decays and $3.5 \times 10^7$ of each muon and tau pairs, with a beam polarization level of 0.8, are given in Table~\ref{tab:chiral}. 

The unpolarized TeraZ cannot access the inclusive polarization asymmetry $A_{\rm LR}$. On the other hand, the $\tau$ forward-backward polarization asymmetry $A_{\rm FB, pol}^{\tau}$ can provide a direct measurement of the electron left-right coupling asymmetry in a clean manner, especially if only the $\tau \rightarrow \pi \nu_\tau$ decay is used (to avoid hadronic uncertainties). An extrapolation from the uncertainties obtained by the ALEPH experiment yields an expected precision $\Delta {\cal A}_{\rm e} = 1.5 \times 10^{-5}$~\cite{Benedikt:2653673} at the FCC-ee TeraZ\footnote{We consider that the bookkeeping theoretical uncertainty of 0.0002 mentioned in Ref.~\cite{ALEPH:2005ab} will have been addressed by then.}. The forward-backward asymmetries for muons and taus are used in combination with this quantity to obtain ${\cal A}_\mu$ and ${\cal A}_\tau$. The forward-backward asymmetry in the $\rm e^+ e^-$ final state could also be used to this effect. The steep dependence of these quantities on the centre-of mass energy is mitigated by the excellent continuous energy calibration of the machine during data taking~\cite{EPOL}. 

From these quantities it is possible to extract {\it (i)} the value of the effective weak mixing angle under the assumption of lepton universality; and {\it (ii)} two ratios of couplings as tests of lepton universality. As suggested in Ref.~\cite{Irles:2019xny}, the ratios of the muon-to-electron and muon-to-tau asymmetries provide a significant cancellation of systematic uncertainties. The results are shown in Table~\ref{tab:chiral}.

\begin{table}[ht]
\begin{center}
\caption{\label{tab:chiral} \small Comparisons between the ILC GigaZ and FCC-ee TeraZ   for the measurements of left-right coupling asymmetries, tests of lepton universality, and measurements of the effective weak mixing angle at the Z pole. Also indicated is the limiting precision on the effective mixing angle from the precision on $\alpha_{\rm QED}(m^2_{\rm Z})$ taking into account for FCC-ee of the improvement on this quantity from the off-peak measurement of the muon forward-backward asymmetry~\cite{Janot:2015gjr}.\vspace{0.3cm}} 
\begin{tabular}{|l|c|c|}
\hline 
Facility  &   ILC-GigaZ  & FCC-ee \\ \hline
Z produced at the peak &  $10^9$  & $4 \times 10^{12}$ \\ \hline
Longitudinal polarization $(P_{\rm e^-}, P_{\rm e^-}) $& $(\pm 0.8, 0.0)$ & $(0.0, 0.0)$ \\ \hline
$\Delta {\cal A}_{\rm e}$ &  $1.2\times 10^{-4}$ & $1.5 \times 10^{-5}$   \\ %\hline
$\Delta {\cal A}_{\mu}$ &  $3 \times 10^{-4}$ & $5 \times 10^{-5}$   \\ %\hline
$\Delta {\cal A}_{\tau}$ &  $3 \times 10^{-4}$ & $5 \times 10^{-5}$   \\ %\hline
$\Delta \frac{  {\cal A}_\mu   }{ {\cal A}_e  } $ &  $1.6 \times 10^{-3}$ & $2.5$ to $4 \times 10^{-4}$   \\ %\hline
$\Delta \frac{  {\cal A}_\mu   } {  {\cal A}_\tau  } $ &  $2.3 \times 10^{-3}$ & $ 3.3 \times 10^{-4}$   \\ \hline
$\Delta \sin^2\theta_{\rm W}^{\rm eff} $ &  $1.5 \times 10^{-5}$ & $ 6 \times 10^{-6}$   \\ %\hline
Hard limit on SM prediction: &&\\
$\Delta \sin^2\theta_{\rm W}^{\rm eff} {\rm from~} \alpha_{\rm QED}(m^2_{\rm Z})  $ &  $1.1 \times 10^{-5}$ & $ 7 \times 10^{-6}$   \\ %\hline
\hline
\end{tabular} 
\end{center}
\end{table}

A similar analysis could be drawn for the b and c quark couplings. The conclusions of this  short analysis are clear: {\it (i)} with the determination of the tau forward-backward polarization asymmetry, all final state left-right coupling asymmetries can be determined at the TeraZ from forward-backward asymmetries measurements for those fermions whose charge can be identified, either on an event-by-event basis or statistically; {\it (ii)} the availability of longitudinal polarization at the ILC GigaZ partly compensates for the much lower luminosity with respect to the FCC-ee TeraZ. {\bf The TeraZ remains nonetheless typically five to ten times more precise for all the measurements of left-right coupling asymmetries}, as shown in Table~\ref{tab:chiral}. 

{\bf For all other measurements, the FCC-ee TeraZ run offers an extremely sensitive program of electroweak, QCD, and flavour physics, as well as numerous searches for rare processes, with 3000 times more statistics than the GigaZ} (Section~\ref{sec:trillions}). In particular, tests of lepton universality in charged current, from tau decays, and in the axial-vector neutral current from the Z leptonic branching ratios, for which mainly statistics matter, can be performed at the TeraZ (resp. GigaZ) at the $10^{-5}$ (resp. few $10^{-4}$) level.  

\subsection{Longitudinal Polarization and Higgs Coupling Determination} 
\label{sec:LongitudinalH}

Another physics argument has recently been proposed~\cite{Bambade:2019fyw}, which is examined below. Reference~\cite{Bambade:2019fyw} claims that, for Higgs coupling determination, {\it beam {\rm [longitudinal]} polarization is a very powerful tool that essentially compensates the advantage of larger event samples claimed  by  the  circular  machines}. This claim has been recently emphasized in several public presentations~\cite{Hitoshi-Yamamoto-FCAL}, where it was stated that  {\it {\rm [longitudinal]} beam polarization is equivalent to a factor of 2.5 in effective luminosity}. A detailed new investigation, however, has led to the same conclusion as above. {\bf Longitudinal polarization brings no information that cannot be obtained otherwise, even in the Higgs sector.}

We note first that the $\kappa$ fits are
insensitive to these subtleties, and essentially independent of whether beams are longitudinally polarized or not. A quick examination of Table~\ref{tab:kappaEFT} shows that the claim of Ref.~\cite{Bambade:2019fyw} is not upheld, even in the EFT framework: the FCC-ee Higgs precisions are significantly better than those of polarized linear colliders. Admittedly, the difference does not entirely account for the factor of 3.3 in luminosity for all couplings, but this is not due to beam polarization. This is because of the combination with HL-LHC projections and, to a lesser extent, to the less sophisticated analyses used by the FCC-ee team. Tables XVIII and XIX of Ref.~~\cite{Bambade:2019fyw} help solve these two issues (as they include no combination with HL-LHC for most couplings, and use the same analyses in both configurations). A summary of these Tables is reproduced in Table~\ref{tab:HiggsPolarization} below.

\begin{table}[ht]
\begin{center}
\caption{\label{tab:HiggsPolarization} \small From Ref.~\cite{Bambade:2019fyw}: Precision on the Higgs boson couplings and total width for the low-energy ILC Higgs factory with 2\,${\rm ab}^{-1}$ at 250\,GeV and with longitudinal beam polarization (middle column), and for a hypothetical ILC configuration with 5\,${\rm ab}^{-1}$ and 1.5\,${\rm ab}^{-1}$ at 250 and 350\,GeV and no beam polarization (right column). The latter also includes precision electroweak measurements foreseen at FCC-ee. An improvement by one order of magnitude of the lepton asymmetry measurement precision with polarized Z$\gamma$ events at 250\,GeV would improve marginally the first four numbers of the middle column -- typically by up to 10\%. The H$\mu\mu$, H$\gamma\gamma$, and HZ$\gamma$ couplings are combined with the HL-LHC projections (which saturate their precision).\vspace{0.4cm}}
\begin{tabular}{|l|c|c|}
%\multicolumn{6}{c}{} \\ 
\hline $\sqrt{s}$ (GeV)  & 250 & 250 + 350 \\ \hline
Lumi (${\rm ab}^{-1}$) &  2 &  5 + 1.5 \\ \hline
Polarization & Yes & No \\ \hline
$g_{\rm HZZ}$ (\%) &  0.57 & 0.34  \\ %\hline
$g_{\rm HWW}$ (\%) &  0.55 & 0.35   \\ %\hline
$g_{\rm Hbb}$ (\%) &  1.0 & 0.62 \\ %\hline
$g_{\rm H\tau\tau}$ (\%) & 1.2 & 0.71 \\ %\hline
$g_{\rm Hcc}$ (\%) &  1.8 & 1.1 \\ %\hline
$g_{\rm Hgg}$ (\%) &  1.6 & 0.96  \\ %\hline
$g_{\rm H\mu\mu}$ (\%) &  4.0 & 3.7 \\ %\hline
$g_{\rm H\gamma\gamma}$ (\%) & 1.1 & 1.0 \\ %\hline
$g_{\rm HZ\gamma}$ (\%) &  9.1 & 8.1 \\ \hline
$\Gamma_{\rm H}$ (\%) & 2.4 & 1.4 \\ \hline
\end{tabular} 
\end{center}
\end{table}

For the couplings not combined with HL-LHC projections, the unpolarized precisions are better than the polarized precisions by factors 1.6 to 1.7 in the EFT framework, which almost entirely accounts for the factor of 3.2 difference in luminosity. This is understood by the fact that both the FCC-ee electroweak precision measurements and the optimized FCC-ee operation model at several centre-of-mass energies independently constrain a number of EFT operators that would be similarly constrained by longitudinally-polarized beams at 250\,GeV. So where does the aforementioned factor of 2.5 in effective luminosity come from? We think that it arises from an over-interpretation of the caption of Fig.~7 in Ref.~\cite{Bambade:2019fyw}, according to which {\it "at 250\,GeV, 2\,$\it ab^{-1}$ with polarized beams yield \underline{comparable} results to a much larger data set of 5\,$\it ab^{-1}$ with unpolarized beams"}. If the FCC-ee operation model did not include a run at the $\rm t\bar t$ threshold, the benefit of polarization would indeed be more visible, but it would still not correspond to a flat increase of the luminosity by a factor of 2.5 at 250\,GeV. For the coupling to weak gauge bosons, the effective luminosity increase would be more like 25 to 60\%\footnote{An improvement by one order of magnitude (with respect to LEP) of the lepton asymmetry precision is required to reach an effective luminosity increase of 60\% for the Higgs coupling to weak bosons.}, and a mere 20\% for the couplings to fermions (b,c, $\tau$) and to gluons. The influence of polarization on the couplings to muons, photons, and Z$\gamma$ is negligible, as the HL-LHC saturates their precision. This marginal benefit of polarization totally fades away in the $\kappa$ fit (even without a run at the top pair threshold) and, in the EFT fit, as soon as a run at 350-365\,GeV is added to the operation model, as optimized for the FCC-ee.  

Over and above that, we note that the $5~{\rm ab}^{-1}$ of FCC-ee$_{240}$ require only 3 years of running, while the $2~{\rm ab}^{-1}$ of ${\rm ILC}_{250}$ require 11.5 to 18.5 years with the baseline parameters (Section~\ref{sec:LowEnergyHiggsFactory}). {\bf All things considered, for Higgs coupling determinations, one year of unpolarized FCC-ee$_{\bf 240}$ with 2 IPs is therefore worth about 6 to 9 years of polarized ${\bf ILC}_{\bf 250}$ for the couplings to weak bosons, and 8 to 12 years for the other accessible couplings.} Improved configurations (4 IPs at the FCC-ee or a doubling of the positron source rate at ILC) would not significantly modify these ratios. 

\section{Will the Accuracy of FCC-ee Higgs Measurements be Affected by Experimental Uncertainties?}
\label{sec:ExperimentalUncertainties}

This question is important because the Higgs-related numbers presented in the previous sections do not include experimental uncertainties. The reason is that, {\bf at FCC-ee, experimental uncertainties are not expected to be a concern for Higgs measurements.} We have found that, for all the measurements we have examined, detailed experimental work -- mostly by collecting and analysing independent information -- will always lead to small experimental uncertainties that decrease as fast as the statistical errors. A few examples are listed below.
\begin{itemize}
    \item The Higgs couplings are extracted from the measurement of the Higgs branching fractions, which in turn depend on the Higgs boson mass. The statistical precision on the Higgs boson mass will be of the order of 10\,MeV at FCC-ee~\cite{Azzi:2012yn}, which translates into a small enough uncertainty on the couplings. Experimentally, the mass determination requires the knowledge of the centre-of-mass energy and, to a lesser extent, of its spread. At $\sqrt{s} = 240$\,GeV, the centre-of-mass energy can be calibrated to 1-2\,MeV with the large samples of Z$\gamma$, WW, and ZZ events, and the method can be itself calibrated at $\sqrt{s} = 160$\,GeV to a few 100\,keV with resonant depolarization. {\rm This possibility is unique to circular colliders.} As for the centre-of-mass energy spread ($\sim 300$\,MeV), it can be measured continuously using ${\rm e^+e^-} \to \mu^+\mu^-$ events~\cite{EPOL}, with a precision of  3\,MeV/$\sqrt{\rm days}$, leading to a negligible impact on the recoil mass uncertainty.   
    \item The efficiencies of the analyses that select Higgs boson events require a perfect alignment of the vertex detector (for flavour tagging efficiencies), and of the tracker (for absolute angle determination), and a perfect calibration of the calorimeters (lepton identification, jet angular resolutions, etc.). The statistical precisions listed in the previous sections require a "standard candle" produced with a sufficiently large rate to provide the necessary alignment and calibration. This standard candle, however, does not really exist at 240\,GeV and above. {\bf A unique possibility of circular colliders is to perform regular runs at the Z pole}, as was done at LEP. At FCC-ee, one hour of data taking will give $3\times 10^8$ Z decays (60 times the whole LEP statistics in each detector) when the collider is in "Higgs mode", and $10^7$ Z decays (twice the LEP statistics in each detector) when the collider is in "top" mode. These samples correspond to about 1000 times the monthly Higgs statistics, and can therefore provide alignment and calibration errors that are negligible with respect to the statistical uncertainty.
    \item The cross section measurement also requires an accurate determination of the integrated luminosity. We have demonstrated that the luminosity can be measured rapidly with low-angle Bhabha events with a precision of 0.01\%~\cite{Ward:2019ooj,LumiPerez}. This is ten times better than the ultimate coupling precision expected from $10^6$ HZ events. 
    \item The experimental magnetic field will not be uniform, and will have significant radial components, because of the compensating solenoid scheme close to the interaction point, aimed at preserving the beam emittance. It will have therefore to be measured accurately in the tracker volume before the tracker installation. The subsequent variations will be followed with the numerous ${\rm e^+e^-} \to \mu^+\mu^-$ events collected in the regular runs at the Z pole (unique at circular colliders), cross-checked with the less numerous, but still abundant ${\rm e^+e^-} \to \mu^+\mu^-$ events at 240\,GeV, and by continuous coil current measurements.
\end{itemize}

So far, we have found no issue that could not be satisfactorily tackled with independent measurements enabled by the large samples offered by FCC-ee. In addition, measurements performed by two or four independent detectors will cross-check each other and lead, intrinsically, to lower overall systematic effects in combined measurements. {\bf In all instances, the limitation was found to be statistics.} 

%{\color{red} 
Another related question came up frequently during the European Strategy symposium in Granada: Are the Higgs projections from FCC-ee too optimistic (use of fast simulations, generator-level estimates, or simple extrapolations)? This question remains an academic debate, as detailed simulations of Higgs production in $\rm e^+e^-$ collisions have been performed for decades in the context of linear collider studies, and have been analysed, and analysed again, in all possible decay channels. The dependence on the collider geometry is, at best, quite subtle. As mentioned in Section~\ref{sec:BeamBackgrounds}, the FCC-ee beam backgrounds, less severe than at linear colliders, have been simulated in all details and have been shown to be negligible for all practical purposes. Besides, the performance of the FCC-ee detectors are currently similar to those used in linear collider studies. A smaller beam pipe, better detector calibration and alignment (as alluded to just above), new analysis techniques, and ten years to develop innovative detectors at up to four interaction points (Section~\ref{sec:TwoInteractionPoints}), may further improve this performance. It is therefore to be expected that FCC-ee analyses yield similar or slightly better selection efficiencies and mass resolutions.

Nonetheless, simple benchmark analyses have been developed in 2012 for most decay channels~\cite{Azzi:2012yn} with a full simulation of the CMS detector, a few of them being cross-checked with a full simulation of the CLIC detector. These detailed simulations have revealed no surprise when compared to linear collider studies. Their outcomes have been taken advantage of to validate two independent dedicated fast simulation software, later used to develop more sophisticated selection algorithms, which in turn confirmed the full simulation estimates. Finally, for those channels not fully analysed by the FCC-ee team, the analysis performance were conservatively extrapolated from the results of full simulations at ILC (for $\sqrt{s} = 240$\,GeV) and CLIC (for $\sqrt{s} = 365$\,GeV). {\bf On the whole, the FCC-ee projections are therefore to be considered conservative.}  
%}

\section{How does a Muon Collider compare (as a Higgs Factory)?}
\label{sec:MuonCollider}

Muons are leptons: circular muon colliders can therefore do, in principle, everything that circular ${\rm e^+e^-}$ colliders offer, in much smaller rings because of the very much reduced synchrotron radiation. Unfortunately, the estimated luminosity at 240\,GeV is typically 100 times smaller for a muon collider than for FCC-ee, making prohibitive the time needed to get a million Higgs bosons.   

On the other hand, muons are heavy: the cross section for $s$-channel Higgs boson production, $\mu^+\mu^- \to {\rm H}$ at $\sqrt{s} = 125$\,GeV, is 40,000 times larger than its ${\rm e^+e^-}$ counterpart, and the reduced synchrotron radiation allows for a superb centre-of-mass energy definition, with a spread comparable to the Higgs boson width of 4.2\,MeV. With a cross section of 18\,pb and an integrated luminosity of 200 to 800\,${\rm pb}^{-1}$ per year, such a muon collider would produce 3500 to 14,000 at the Higgs pole, and would allow a scan of the Higgs resonance -- similar to what LEP did with the Z boson -- for a direct determination of the Higgs boson mass and width and some branching ratios. 

It was shown in Ref.~\cite{fcc-ee-projections} that, in less than a decade and with an integrated luminosity of $5\,{\rm fb}^{-1}$, the Higgs boson mass would be measured with an exquisite precision of 0.1\,MeV, while the precision on the Higgs boson total width would be of the order of 0.25\,MeV (corresponding to a relative precision of 6\%). In addition, a few branching ratio products (BR$_{\mu\mu}$BR$_{\rm vis}$, BR$_{\mu\mu}$BR$_{\rm bb}$, BR$_{\mu\mu}$BR$_{\rm WW}$, BR$_{\mu\mu}$BR$_{\rm \tau\tau}$) can be measured with precision of 4\%, 2.5\%, 3\% and 10\%, respectively. The other branching fractions would need a very significantly larger luminosity to be measured with a meaningful accuracy. A $\kappa$ fit, performed with the assumption of no Higgs boson exotic decays (invisible or not) and by fixing non-measureable couplings to their SM values, leads to the results presented in Table~\ref{tab:MuonCollider}, which are compared with the FCC-ee projections. Except for the Higgs boson mass and $g_{\rm H\mu\mu}$, the experimental precisions on Higgs parameters fall short of those of a dedicated ${\rm e^+e^-}$ collider. Moreover, circular muon collider designs are much less well developed and will probably require several decades of R{\&}D to ascertain their feasibility.

\begin{table}[htbp]
\begin{center}
\caption{\label{tab:MuonCollider} \small Precision on the Higgs boson couplings and on the Higgs boson mass and width for the  low-energy muon collider Higgs factories at $\sqrt{s}= 125\,{\rm GeV}$, compared to FCC-ee~\cite{fcc-ee-projections}. The muon collider numbers assume that the non-measured couplings and the invisible Higgs boson decays are fixed to their Standard Model values, and assume no BSM decays. No combination with HL-LHC and no theory uncertainties are included here.\vspace{0.4cm}}
\begin{tabular}{|l|c|c|}
%\multicolumn{6}{c}{} \\ 
\hline Collider & {\small $\mu$Coll$_{125}$} & FCC-ee$_{240\to 365}$ \\ \hline
Lumi (${\rm ab}^{-1}$) &  0.005 &  5 + 0.2 + 1.5 \\ \hline
Years & 6 to 10 & 3 + 1 + 4 \\ \hline
$g_{\rm HZZ}$ (\%) &  SM  & 0.17  \\ %\hline
$g_{\rm HWW}$ (\%) &  3.9 & 0.43   \\ %\hline
$g_{\rm Hbb}$ (\%) &  3.8 & 0.61 \\ %\hline
$g_{\rm Hcc}$ (\%) &  SM & 1.21 \\ %\hline
$g_{\rm Hgg}$ (\%) &  SM & 1.01  \\ %\hline
$g_{\rm H\tau\tau}$ (\%) & 6.2 & 0.74 \\ %\hline
$g_{\rm H\mu\mu}$ (\%) &  3.6 & 9.0 \\ %\hline
$g_{\rm H\gamma\gamma}$ (\%) &  SM & 3.9 \\ \hline
$\Gamma_{\rm H}$ (\%) &  6.1 & 1.3  \\ 
$m_{\rm H}$ (MeV) &  0.1 & 10.  \\ \hline
BR$_{\rm inv}$ (\%) &  {\small SM} & 0.19 \\ 
BR$_{\rm EXO}$ (\%) &  {\small SM} & 1.0 \\ \hline
\end{tabular} 
\end{center}
\end{table}

{\bf A muon collider at $\sqrt{s} = 125$\,GeV is an elegant Higgs factory concept, but not yet competitive in precision and indirect sensitivity to new physics.} The muon collider technology, however, might be the only way forward to get to multi-TeV lepton colliders (Section~\ref{sec:DeadEnd}).  

\section{Can I do more than Higgs Physics at FCC-ee?}
\label{sec:MuchMore}

As mentioned in Section~\ref{sec:LowEnergyHiggsFactory}, {\bf FCC-ee is not only an excellent Higgs factory: it is also the ultimate electroweak and flavour factory}. The FCC-ee is a general-purpose precision instrument for the continued in-depth exploration of nature at the smallest scale, on all fronts~\cite{Benedikt:2651299,Mangano:2651294}. It is optimized to study electroweak and strong interactions with high precision~\cite{Beate,Jorgen} in the production and decay of the four heaviest particles of the Standard Model, with samples of $5\times 10^{12}$ Z bosons, $10^8$ W pairs, $10^6$ Higgs bosons, and $10^6$ top-quark pairs, acquired in a clean ${\rm e^+e^-}$ collision experimental environment (Table~\ref{tab:OperationModel}). Precision flavour measurements~\cite{Zoccoli} of $\tau$ leptons and b- and c-quarks from Z decays (with samples of several $10^{11}$ to several $10^{12}$ of these particles), and precision studies of gluons from Higgs decays (with 100,000 ${\rm H \to gg}$ events), are also on offer.

The obvious next question is {\bf why are yet more precision measurements needed?} A first, agnostic, answer to this question was already given in the conclusions of the previous strategy document from the CERN Council in 2013~\cite{ESPP2013}: {\it ``There  is  a  strong  scientific  case  for  an  electron-positron  collider,  complementary  to  the  LHC, that  can  study  the  properties  of  the  Higgs  boson  and other particles with unprecedented precision and  whose  energy  can  be  upgraded."}. {\bf The FCC-ee fills the bill perfectly.}

How is this precision useful? After all, the discovery of the Higgs boson, with a mass of 125\,GeV, has completed the matrix of particles and interactions that constitute the Standard Model, as developed over several decades. This model is a consistent and predictive theory, which has so far proven successful at describing all phenomena accessible to collider experiments. So why test the Standard Model with one or two orders of magnitude greater precision? Several experimental facts require the extension of the Standard Model and explanations are needed for observations such as the abundance of matter over antimatter~\cite{Zoccoli}, the striking evidence for dark matter~\cite{Marcela}, and the non-zero neutrino masses~\cite{Zito}. A number of theoretical issues also  need to be addressed experimentally, including the hierarchy problem, % the neutrality of the Universe,
the stability of the Higgs boson mass under quantum corrections, unification of the fundamental interactions and the strong CP problem.

Possible answers to these open questions seem to require the existence of new particles and phenomena over an immense range of mass scales and coupling strengths, which could have masses too large or couplings too small to be observed at the LHC~\cite{Marcela,Zito,Paris}. To make things more challenging, it is worth recalling that the predictions of the top quark and Higgs boson masses from a wealth of precision measurements, collected in particular at ${\rm e^+e^-}$ colliders and from other precise low-energy experimental inputs, were made strictly within the Standard Model framework, under the assumption that there is no new physics beyond it. We must, therefore, look for well-hidden new physics that does not significantly modify the quantum corrections upon which the Standard Model predictions were made. 

History has shown that the existence, properties and approximate mass values of heavier Standard Model particles (Z, W, Higgs, and top) were picted before their actual observation on the basis of a long history of experiments and theoretical interpretation. In this context, a decisive improvement in precision measurements of electroweak observables and of particle masses would play a crucial role, providing sensitivity to a large range of new physics possibilities~\cite{David:2015waa}. The observation of significant deviation(s) from the Standard Model predictions would definitely be a discovery. The prospects for such a discovery depend upon a significant improvement in experimental and theoretical precision. They also require the largest possible set of measured observables to eliminate spurious deviations, and most importantly to possibly reveal a pattern that would guide the theoretical interpretation and point to the source and the scale of new physics. Similarly, the search for new particles with extremely small couplings or for forbidden phenomena, in Z or Higgs boson decays in particular, could give the first clues towards the understanding of some of the remaining fundamental questions.

{\bf Improved precision on all electroweak fronts increases the discovery potential.} A lepton collider with the highest luminosities at centre-of-mass energies between $\sim 90$ and $\sim 400$\,GeV has the strongest physics case in this respect, as it would cover the Z pole, the W- and top-pair production thresholds, and allows for copious Higgs boson production. With such a device, precision electroweak physics, precision Higgs physics, and measurements of the top quark and W boson properties will give orders of magnitude improvements. High-energy physics requires the next lepton collider to be an electroweak, flavour, and Higgs factory at the precision frontier, capable of providing a long list of exquisite precision measurements probing new physics and complementing each other. 

\begin{figure}[htbp]
\centering
\begin{sideways}
\begin{minipage}{23.5cm}
\vspace{-0.5cm}
\caption{\label{fig:FCCPhysics} \small Palette of physics measurements/topics offered in $\rm e^+e^-$ and proton-proton collisions as a function of the centre-of-mass energy. Green cells indicate the measurements and discovery potential unique to FCC. Blue (pink) cells illustrate the topics where FCC (CLIC) provides the best prospects either in measurement precision or high-mass/feebly-coupled particle discovery reach. The text in these cells spells out the key ingredient for the collider to lead the corresponding topic. (*) means that FCC-hh measurements need to be combined with FCC-ee measurements to reach the quoted precision. Precision electroweak measurements at high energy~\cite{Farina:2016rws} lead to similar sensitivity to nearby new physics for CLIC and FCC-hh (a slightly better measurement of the electroweak parameter $Y$ with CLIC, slightly better $W$ and $Z$ determinations with FCC-hh). For direct searches at high masses, the FCC-hh provides a general new physics reach to the highest masses, while CLIC offers an easier closure of supersymmetric parameter space for electroweakinos below 1.5\,TeV and with small mass difference with the lightest supersymmetric particle (indicated by the hatched pattern).\vspace{0.3cm}}
\includegraphics[width=\columnwidth,angle=0]{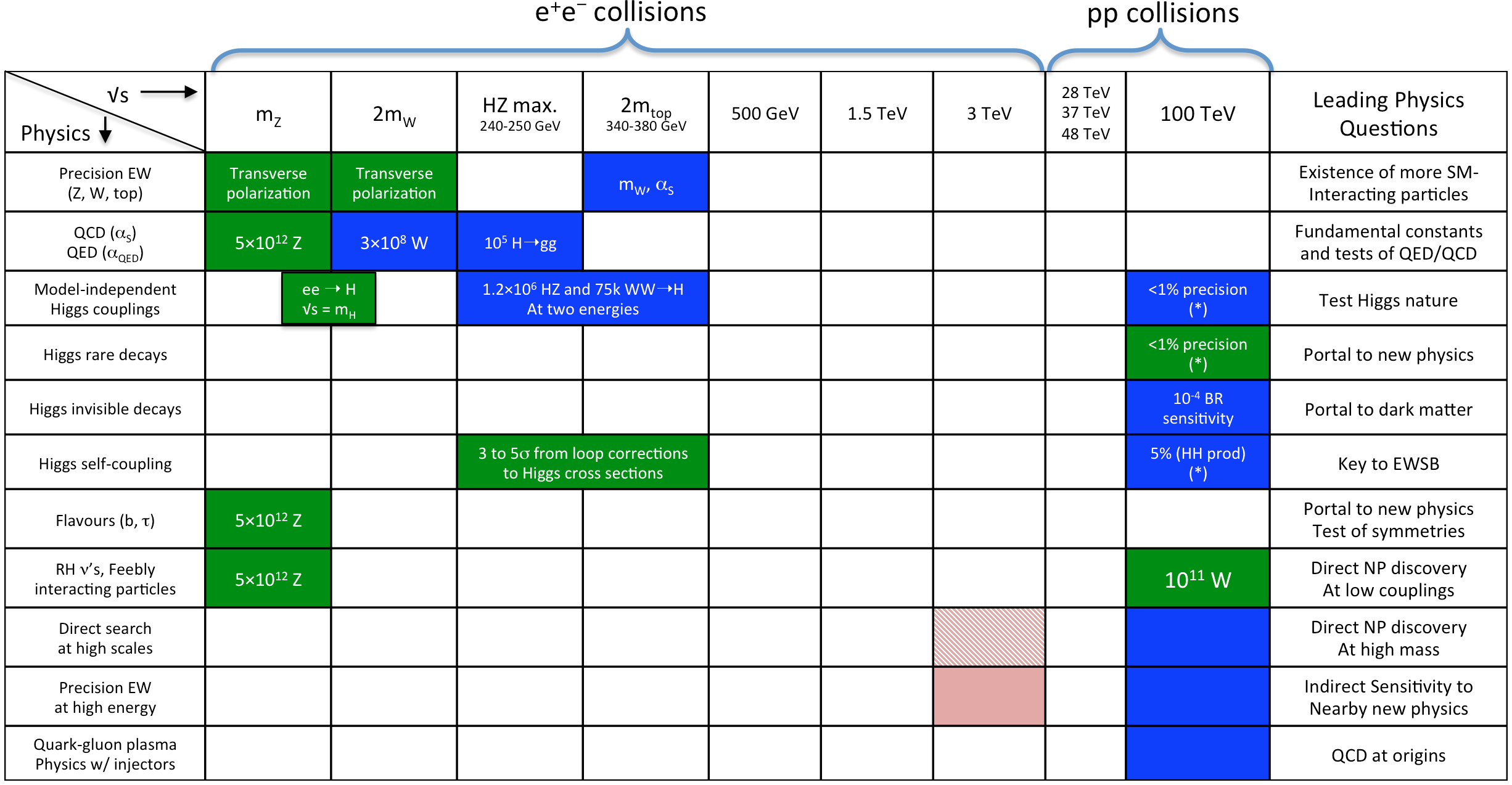}
\end{minipage}
\end{sideways}
\end{figure}

{\bf Again, FCC-ee fills the bill ideally.} It is the most powerful of all the proposed $\rm e^+e^-$ colliders at the electroweak scale, giving access to much higher scales and/or much smaller couplings. The integrated FCC programme, with a synergistic combination of the measurements from FCC-ee and FCC-hh, uniquely maps the properties of the Higgs boson (which also greatly benefit from the precision electroweak measurements at FCC-ee). The FCC-hh also improves by close to an order of magnitude the discovery reach for new particles at the highest masses, with respect to HL-LHC. The physics menu offered by the FCC integrated programme, as summarized in the Granada European Strategy symposium~\cite{Beate,Jorgen,Zoccoli,Marcela,Zito,Paris}, is illustrated in the table shown in Fig.~\ref{fig:FCCPhysics} and compared to that of other collider projects.

\section{Why do we need At Least \texorpdfstring{$\bf 5\times 10^{12}$}{trillions} Z Decays?}
\label{sec:trillions} 

Figure~\ref{fig:FCCPhysics} shows that the extremely high FCC-ee luminosity at and around the Z pole is a key ingredient in at least three physics topics that are unique to FCC-ee, namely high-precision electroweak tests, high-precision flavour physics and rare decays, and searches for less-than-weakly- (feebly-)\,interacting particles. These examples of the need for very high statistics are discussed in turn below.

The global consistency check of the SM through electroweak precision observables at the Z pole and beyond (together with measurements of W-boson, Higgs-boson, and top-quark properties) is one of the pillars of our current understanding of the electroweak scale. The high luminosity of FCC-ee will deliver a total of about $5\times 10^{12}$ Z decays to the detectors at and around the Z pole in four years of operation, taking these tests to the next level and providing unique probes of physics beyond the SM. In particular, FCC-ee will enable a measurement of the weak mixing angle with a statistical precision of $3\times 10^{-6}$, which is well matched to the experimental systematic uncertainty of $2$--$5 \times 10^{-6}$~\cite{Mangano:2651294}. The completion of this electroweak precision programme will need an integrated luminosity of at least $30\,{\rm ab}^{-1}$ just below ($\sqrt{s} \simeq 88$\,GeV) and just above ($\sqrt{s} \simeq 94$\,GeV) the Z pole, to be accumulated in less than two years of operation, so as to obtain a statistically-limited three- to four-fold improvement of the measurement of the electromagnetic coupling constant, $\alpha_{\rm QED}(m_{\rm Z}^2)$~\cite{Janot:2015gjr}. The current precision on this parameter would otherwise limit the ability of the precision electroweak tests to corner the energy scale at which new phenomena may occur. 

A second pillar of the SM is the determination of the profile of the Cabibbbo-Kobayashi-Maskawa (CKM) quark-mixing matrix to probe the CKM paradigm~\cite{Kobayashi:1973fv}. An effective parame\-te\-ri\-zation of BSM contributions~\cite{Soares:1992xi} can be used to infer the BSM scale in neutral meson mixing processes. The anticipated landscape of the future experimental programme is rich,  with the advent of the SuperKEK\,B-factory and the Belle\,II experiment, and of the ongoing and foreseen LHCb upgrades. There is a remarkable complementarity between the two experiments. The ultra-high statistics expected at the LHCb Phase II experiment (HL-LHC) will be instrumental for measuring the CKM $\gamma$ angle to sub-degree precision, and the flavour-tagging versatility of the asymmetric B factory will yield the same performance for the $\beta$ angle, to cite two of the most important observables of the consistency test.  An order of magnitude improvement in precision can be expected, which translates into a factor four for the energy scale~\cite{Charles:2013aka}.  A total of about $10^{12}$ $\rm b \bar b$ pairs, available with a sample of $5 \times 10^{12}$ Z decays, will challenge the precisions of these measurements, and push forward the search for unobserved phenomena such as CP-symmetry breaking in the mixing of beautiful neutral mesons~\cite{Benedikt:2651299}. In parallel, searches for rare decays will make FCC-ee a discovery machine. Lepton-flavour-violating (LFV) Z decays, rare and LFV $\tau$ decays, searches for heavy neutral leptons and rare b-hadron decays have all been explored in Ref.~\cite{Benedikt:2651299} as benchmark or flagship searches, illustrative of the unique potential of a high-luminosity Z factory. The precision of the measurements relies on the vertexing capabilities of the experiments to take benefit of the boosted topologies at the Z energy, but most are limited in precision by the statistical size of the sample.  A minimum of $5 \times 10^{12}$  Z decays has been shown to be necessary to make, for example, a comprehensive study of the rare electroweak penguin transitions $\rm b \to s \tau^+ \tau^-$~\cite{Kamenik:2017ghi}. For example, about 1000 events with a reconstructed $\rm \overline{B}^0 \to K^{\ast 0} \tau^+\tau^-$ are expected in such a sample. Should the current ``flavour anomalies"~\cite{Graverini:2018riw} persist, the study of b-hadron decays involving $\tau$'s in the final state will be required to sort out possible BSM scenarios. If these flavour anomalies do not survive, the study of Z couplings to third-generation quarks and leptons still constitute an excellent opportunity to unravel BSM physics. 

Finally, the FCC Physics CDR~\cite{Mangano:2651294} has documented the extraordinary sensitivity to feebly (less-than-weakly) coupled  particles,  ranging  from  right-handed neutrinos down to the see-saw limit in a part of parameter space favourable for generating the baryon asymmetry of the Universe, to axions and dark photons. Feebly-interacting particles are well-motivated in many extensions of the SM~\cite{Hewett:2012ns,Essig:2013lka,Alexander:2016aln}, but it is hard to predict their properties (couplings and masses), leading to the so-called Log-crisis~\cite{Gilad}. For such elusive particles, each order of magnitude increase in luminosity at the Z pole can be put to good use. Heavy right-handed neutrinos, denoted by N in the following, are an excellent illustrative example. The combination of electroweak precision and lepton universality tests (Section~\ref{sec:LongitudinalZ}) with $5\times 10^{12}$ Z allows upper limits to be set on the square of the right-handed neutrino mixing angle $\Theta$ with their left-handed counterparts at the $10^{-5}$ level, for masses $m_{\rm N}$ that extend well beyond 100\,TeV. For $m_{\rm N}$ below 100\,GeV, direct, background-free, searches for $\rm Z \to \nu N$ with long-lived N become possible. The large Z sample promised at FCC-ee allows sensitivities to $\Theta^2$ to reach values down to $10^{−11}$. In some regions of the ($m_{\rm N},\Theta^2$) parameter space, several hundred signal events are expected to be observed, which would allow a first determination of the mass and lifetime of the right-handed neutrino and establish its relative decay rate into the three lepton flavours. Both indirect and direct sensitivities would improve with larger statistics. 

{\bf In conclusion, the electroweak precision tests, the KM consistency checks and flavour physics in general, including the study of rare decays, and the search for feebly-interacting particles, all require at least $\bf 5 \times 10^{12}$  Z decays %-- and would greatly benefit from even higher statistics -- 
in order to match the necessary precision as indicated by the global fits.} We have only begun to investigate the physics offered by such a TeraZ factory, and much more will emerge when a detailed and systematic study is carried out. 

\section{Why is FCC-ee More Precise for Electroweak Measurements?}

The exceptional precision of FCC-ee comes from several features of the programme:
\begin{itemize}
    \item Extremely high statistics of $5\times10^{12}$ Z decays, $10^8$ WW, $10^6$ ZH, and $10^6$ ${\rm t\bar t}$ events;       
    \item  High-precision (better than 100\,keV) absolute determination of the centre-of mass energies at the Z pole and WW threshold, thanks to the availability of transverse polarization and resonant depolarization~\cite{EPOL}. This is a unique feature of the circular lepton colliders, ${\rm e^+ e^-}$ and $\mu^+ \mu^-$. At higher energy, WW, ZZ and Z$\gamma$ production can be used to constrain the centre-of-mass energy with a precision of 2 and 5\,MeV, at the ZH cross-section maximum and at the ${\rm t\bar t}$ threshold respectively. At all energies, ${\rm e^+e^-}\to \mu^+\mu^-$ events, which at the Z pole occur at a rate in excess of 3 kHz, provide by themselves in matters of minutes the determination of the centre-of mass energy spread, the residual difference between the energies of ${\rm e^+}$ and ${\rm e^-}$ beams and a (relative) centre-of-mass energy monitoring with a precision that is more than sufficient for the precision needs of the programme.   
    \item The clean environmental conditions and an optimized run plan allow a complete programme of ancillary measurements of presently input quantities that currently limit the precision of precision EW tests. This is the case of the top quark mass from the scan of the t\={t} production threshold; of the unique, direct, measurement of the QED running coupling constant at the Z mass from the Z-$\gamma$ interference; of the strong coupling constant by measurements of the hadronic to leptonic branching fractions of the Z, the W and the $\tau$ lepton; and of the Higgs and Z masses themselves. \\
\end{itemize}

If future theory uncertainties match the FCC-ee experimental precision (Section~\ref{sec:TheoryErrors}), the many different measurements from FCC-ee will provide the capability to exhibit and decipher signs of new physics. Here are two examples: the EFT analysis searching for signs of heavy physics with SM couplings shows the potential for exhibiting signs of new particles up to around 70\,TeV; with a very different but characteristic pattern, observables involving neutrinos would show a significant deviation if these neutrinos were mixed with a heavy counterpart at the level of one part in 100,000, even if those were too heavy to be directly produced. 

\section{Will Theory be Sufficiently Precise to Match this Experimental Precision?}
\label{sec:TheoryErrors}

When this question was asked in 2013~\cite{Gomez-Ceballos:2013zzn}, the prospects of precision electroweak measurements were at the level of 100\,keV or better for the Z mass and width, 500\,keV for the W mass, $ \pm 5 \times 10^{-6}$ for ${\rm \sin^2{\theta_{\rm w}^{\rm eff}}}$, $\pm 0.00015$ for $\rm \alpha_s (m_{\rm Z}^2) $, etc. {At the time, the precision of the theoretical prediction for all these observables was nowhere near the required level; for example, the Z width was calculated for LEP with an accuracy of 0.5\,MeV, an order of magnitude larger than what is needed for FCC-ee, challenging the theory community. The required precision at the Z pole indeed demands calculations of quantum electroweak corrections, with all SM heavy particles included, at the three-loop level, i.e., with hundreds of thousands of Feynman diagrams of higher complexity.}

{The theory community is now more confident of rising to the challenge,  based on breakthroughs using new numerical methods for multi-loop calculations: the {\it sector decomposition method}, widely used for LHC predictions; and the {\it Mellin-Barnes representation} of Feynman integrals. These methods have been used to complete successfully two-loop electroweak calculations at the Z pole~\cite{Dubovyk:2016aqv,Dubovyk:2018rlg}. Other numerical or analytical approaches are also developing, based, e.g., on calculations in $d=4$, on unitarity/loop-tree duality methods, or on other fresh ideas, as reviewed in Refs.~\cite{Blondel:2018mad,Blondel:2019vdq}. The quantum field theory needed for FCC-ee is deeply intertwined with
contemporary cutting-edge mathematical studies, requiring even newer concepts and innovations. From this perspective, the experimental precision envisioned at FCC-ee offers a very attractive challenge to the next generation of young theorists for the coming decades.}

{For this reason}, a series of FCC-ee workshops~\cite{mini0,mini-pbp, Azzi:2017iih, mini, Blondel:2018mad, mini2019} has gathered a rapidly-growing community of experts in precision calculations. The conclusions of the 2018 workshop stated: 
{\bf "We anticipate that, at the beginning of the FCC-ee campaign of precision measurements, the theory will be precise enough not to limit their physics interpretation.  This statement is however conditional to sufficiently strong support by  the physics community and the funding agencies, including strong training programmes"}. The growth in interest in the theory community has been impressive, with successive reports attracting ever more authors from more institutes.  

The ability of the theory community to meet the challenges posed by the FCC-ee electroweak precision programme is supported by the remarkable progress made in developing new calculational techniques and the challenging higher-order calculations for hadronic processes. Results such as the N$^3$LO cross section for Higgs production and Drell-Yan processes, and the NNLO results for most 2$\to$2 processes, were unthinkable just few years ago, and have been stimulated by the concrete needs of the LHC programme. Similarly, the perspective of the FCC-ee precision measurements will stimulate huge progress in the particle physics theory community. 

A preliminary evaluation of the required support~\cite{Blondel:2018mad} concluded that electroweak three-loop calculations and the leading four-loop electroweak-QCD corrections are needed for electroweak observables at the Z pole, and at least two-loop corrections for the double-resonant WW production. {While being a serious challenge, it is also recognized as a fascinating opportunity for the theory community.} {\bf Matching the experimental precision across the board has been estimated to require a dedicated effort of 500 person-years. This requirement has been listed as a strategic, high-priority, item in the FCC-ee CDR.  This challenge will have to start with a strong training programme for the younger generation.}

{As a corollary, sophisticated Monte Carlo event generators, including the aforementioned multi-loop electroweak and QCD corrections and based on soft-photon resummation, will have to be developed~\cite{Blondel:2018mad,Blondel:2019vdq}. Indeed, none of the legacy codes used at LEP provide the precision necessary for FCC-ee, especially in the Z pole and WW threshold stages; and the significant advances in QCD codes for LHC are of limited use for FCC-ee. Moreover, for validation purposes, at least two different codes will have to be built independently for each (class of) final states, as analytic calculations will no longer be possible with the precision required by FCC-ee. Last but not least, the role of Monte Carlo event generators will be even more prominent than at LEP in the extraction of (re-defined) pseudo-observables from the data. These additional challenges must be tackled with no delay.} 

\section{What can be discovered at FCC-ee?}

%{\it Discovery, n: The action of finding out or becoming aware of something for the first time; the action of being the first to find (a place); the action of bringing to light something (as a scientific phenomenon) which was previously unknown.}~\cite{OED}

As the most powerful of all ${\rm e^+e^-}$ collider projects at the electroweak scale, FCC-ee proposes, with centre-of-mass energies from 88 to 365\,GeV and in a coherent research programme of about 15 years, multifaceted exploration to maximise opportunities for major discoveries. With 20 to 50-fold improved precision in all electroweak observables, and with up to 10-fold more precise and model-independent determinations of the Higgs couplings and width  in a complementary way, FCC-ee, despite its limited centre-of-mass energy, probes new physics effects at scales as high as 10 to 100\,TeV. New physics discovery will come from the fact that the Standard Model does not fit. Specific deviation patterns among these measurements would offer a tool to select the compatible  underlying theory(ies).  

Furthermore, the high-statistics samples delivered by FCC-ee offer unique opportunities of direct discoveries beyond the precision programme. Other signals of new physics could arise from the observation of small flavour-changing neutral currents or lepton-flavour-violating decays, from the observation of dark matter in Z and Higgs invisible decays, or by the direct discovery of particles with extremely weak couplings in the 5 to 100\,GeV mass range, such as right-handed neutrinos and other exotic particles (see FCC physics CDR~\cite{Mangano:2651294}, Section 12.4 and Chapter 13). These scenarios~\cite{Antusch:2016ejd,Bauer:2018uxu} are well-motivated and, despite the low masses involved, consistent with the constraints imposed by precision measurements. 

Finally, a circular collider such as FCC-ee offers a unique advantage. Just as it was for LEP and LHC, it is a long-term investment. Hadron colliders remain the best way to reach high energy parton-parton collisions with high luminosity for the foreseeable future. In the integrated FCC plan, the FCC-ee tunnel is designed from the outset to host subsequently a hadron collider. The FCC-hh, with a centre-of-mass energy of 100\,TeV and able to accumulate $30\,{\rm ab}^{-1}$ in 25 years, is designed for the direct discovery of particles with Standard Model couplings and with masses of up to 40\,TeV, as illustrated in Fig.~5 of Ref.~\cite{Benedikt:2653673}. Direct searches at the FCC-hh will of course be further motivated and possibly guided, if a discovery made at the lepton collider, in precision measurements or otherwise, points to the existence of new particle(s) coupling to the SM particles, with specific signature or patterns of deviations depending on its underlying nature. The FCC-hh is also a remarkable factory of $10^{10}$ Higgs bosons and more than $10^{13}$ W bosons, allowing a set of complementary and synergistic measurements and searches for rare or forbidden decays. Only FCC-ee provides the necessary platform to achieve this, through its evolution into the FCC-hh programme. The FCC-hh discovery reach at the energy and precision frontiers, when combined with the FCC-ee precision  and sensitivity, exceeds that of any proposed linear collider energy upgrade, on any foreseeable timescale (Fig.~\ref{fig:FCCPhysics}).

\section{Is the FCC-ee Project "Ready to Go"?}
\label{sec:ReadyToGo}

{\bf All the technologies required for the  first phase of the FCC programme, namely FCC-ee, are at hand and the design proposed in the FCC-ee CDR can be built now.} In several places, additional optimization and  detailed design are planned, with the aim of further reducing the cost and the risks, of improving the baseline performance, or of finalizing a number of choices (such as in the design of the injector). 

The main technological components of the collider are its radiofrequency (RF) system and, in the arcs, the magnets and the vacuum system. Research and development  on higher-efficiency RF power sources (klystrons, inductive output tubes, 
or solid-state amplifiers) can boost the overall energy efficiency. The RF cavities are based on thin-film technology.
The design foresees, for the HZ working point, operation of Nb/Cu cavities at twice the field gradient achieved for LEP2, a goal that appears within close reach. Better performance might be attained with more advanced technologies, e.g., Nb$_{3}$Sn film on Cu, for the ${\rm t\bar t}$ working point (for which bulk Nb cavities are currently foreseen). For the arcs, low-power twin-aperture dipole and quadrupole magnets have been developed and prototyped. 

In the injector complex, an extension of the 6\,GeV Linac to 20\,GeV could avoid the use of a modified SPS as a pre-booster ring, and would offer greater flexibility for the filling of the collider.  For the positron source, both a conventional target and a hybrid target consisting of a crystal followed by an amorphous material are under study. The design of the top-up booster can also be further optimized with regard to cost.

Several reviews have emphasized that a technical schedule would probably be significantly shorter than that which is proposed. The present schedule includes time for authorizations and fundraising, as well as the time to allow for completion of the full HL-LHC programme (Section~\ref{sec:HLLHC}). 

\section{What is the cost of the FCC-ee?}
\label{sec:cost}

\subsection{What are the FCC-ee Construction Costs?}

The estimated costs of the  FCC were shown in the public presentation of the FCC-CDR on 4--5 March 2019~\cite{FCC-CDR-public-benedikt2}.  The breakdown of cost of the full FCC programme (ee and hh) is as follows:
\begin{enumerate}
    \item 4 GCHF for the FCC-ee collider and injector;
    \item 17 GCHF for the FCC-hh collider and injector (of which 9.4 GCHF for the magnets);
    \item 7.6 GCHF for the common civil engineering and technical infrastructure.
\end{enumerate}
If one aims at reaching one day 100\,TeV pp collisions, these estimates place the FCC-ee cost in the same ball park as the ILC$_{250}$ proposed to the Japanese government (6.5 GCHF, without the site construction) and the first stage of CLIC (of the order of 6 GCHF). We note that these are target costs, and that the additional time given to the hadron collider by the FCC-ee programme may well result, by virtue of a longer R\&D period, in some combination of lower cost and higher performance, that could be set off against the FCC-ee cost. The physics case of the FCC-ee, however, justifies the initial investment in its own right.

\subsection{What are the Costs of Operating FCC-ee?}

The total electrical energy consumption over the fourteen years of the FCC-ee research programme is estimated to be around 27 TWh~\cite{Benedikt:2653669},  corresponding to  an average electricity consumption of 1.9 TWh/year over the entire operation programme, to be compared with  the 1.2 TWh/year consumed by CERN today and the expected 1.4 TWh/year for HL-LHC\footnote{For comparison, the LEP2 energy consumption ranged  between 0.9 and 1.1 TWh/year.}. At the CERN electricity prices from 2014/15, the electricity cost for FCC-ee collider operation would be about 85 MEuro per year. In the HZ running mode, about one million Higgs bosons are expected to be produced in three years, which sets the price of each FCC-ee Higgs boson at 255 Euros. A similar exercise can be done for the first stage of CLIC, expected to consume 0.8 TWh/year over 8 years at 380\,GeV to produce about 150,000 Higgs bosons, which sets the price of a CLIC Higgs boson at about 2000 Euros. Finally, with the official ILC operation cost in Japan of 330\,MEuro per year~\cite{Bambade:2019fyw}, its {11.5 to 18.5 years} of operation (Section~\ref{sec:LowEnergyHiggsFactory}), and the 500,000 Higgs bosons produced in total, the price of an ILC Higgs boson is between {7,000 and 12,000 Euros}, i.e., between {30 and 50 times} more expensive than at FCC-ee. These operation costs are summarized in Table~\ref{tab:OperationCost}.

\begin{table}[htbp]
\begin{center}
\caption{\label{tab:OperationCost} Operation costs of low-energy Higgs factories, expressed in Euros per Higgs boson.}\vspace{0.4cm}
\begin{tabular}{|c|c|c|c|}
\hline
Collider              & ILC$_{250}$ & CLIC$_{380}$ & FCC-ee$_{240}$\\ \hline
Cost (Euros/Higgs)     & {7,000 to 12,000} & 2,000 & 255 \\ \hline
\end{tabular}
\end{center}
\end{table}

\section{Can FCC-ee be the First Stepping Stone for the Future of our Field?}
\label{sec:DeadEnd}

This question is key in the choice for the next facility. 
It is often argued that high-energy physics needs an ${\rm e^+e^-}$  Higgs factory; that it should preferably be built with a technology that is important for the future of the field; that circular ${\rm e^+e^-}$ colliders are limited in centre-of-mass energy because of synchrotron radiation; that therefore linear colliders are the only way to go to higher energies, and hence the next machine should be a linear collider. However, there are strong counterarguments.

\subsection{Is a linear collider the best ``Electroweak and Higgs Factory" that can be built?} 

Figure~\ref{fig:OperationModel} gives the answer for a low-energy Higgs Factory: circular colliders are an order-of-magnitude superior in luminosity than the linear colliders that are on the table, namely CLIC and ILC. The cross-over in luminosity occurs at a centre-of-mass energy around 400\,GeV. All the other particles of the Standard Model, and specifically the four heaviest ones, the Z, the W, the Higgs and the top, can be produced below this energy.
Besides, the availability of two (and perhaps four) collision points, of exquisite beam energy calibration to one part per  million, and the huge luminosities at the Z pole and the WW threshold, are all in favour of a circular machine. Control of the ${\rm e^{\pm}}$ beam helicity is very interesting and convenient but, in most cases, equivalent information can  be obtained by other means, such as final-state polarization measurements or angular distributions, as discussed above (Section~\ref{sec:Polarization}).
High-energy ${\rm e^+e^-}$ linear colliders (reaching $\gg$ 500\,GeV) can measure directly the top Yukawa coupling and the Higgs self-coupling. As already mentioned in Section~\ref{sec:EnergyUpgrades}, however, the ttH coupling will be more precisely measured already at the HL-LHC, and the Higgs self-coupling can be better measured at the 100\,TeV FCC-hh than at any of the proposed linear colliders, thanks to the much greater statistics% and provided the precision measurements from the earlier ${\rm e^+e^-}$ collider are available
. Other measurements, and especially the precision electroweak observables, are all measured better at a circular ${\rm e^+e^-}$ collider. {\bf In summary, the FCC (ee and hh, with an eh option) is the best Electroweak and Higgs Factory on the market}. For several reasons, an $s$-channel muon collider is not competitive (Section~\ref{sec:MuonCollider} and Refs.~\cite{fcc-ee-projections,BlondelAries2018}).

\subsection{Can one build a long-term strategy based on linear \texorpdfstring{${\bf e^+e^-}$}{ee} colliders?} 

It is a good feature that the linear collider energy can be upgraded in steps. One should be aware, however, that these upgrades are not cheap and will take longer than the "technically feasible" times usually advertised. An upgrade of ILC from 250 to 500\,GeV would raise the cost  by a factor of 1.6, and a larger factor would be needed to reach 1\,TeV, making it a $\sim 20$ billion machine. This is also the cost announced for CLIC, when upgraded to 3\,TeV. An interesting set of scaling laws including cost estimates can be found in the talk by Vladimir Shiltsev at the EPS-HEP conference in Vienna 2015~\cite{ShiltsevVienna,Shiltsev:2014cva}, or in a recent compilation by J.-P.~Delahaye~\cite{DelahayeAries2018}. {\bf In the case of CERN, it is difficult to envision how a facility delivering beams to only one experiment, and with a centre-of-mass energy limited to 3\,TeV, could provide the future of the laboratory, at least not in the present scientific context}, unless a new particle that could be copiously produced in lepton collisions in this energy range were to be discovered at the LHC.

\subsection{Can one go beyond \texorpdfstring{3\,TeV}{3TeV} in lepton collisions?}
\begin{figure}[htbp]
\centering
\includegraphics[width=0.50\textwidth]{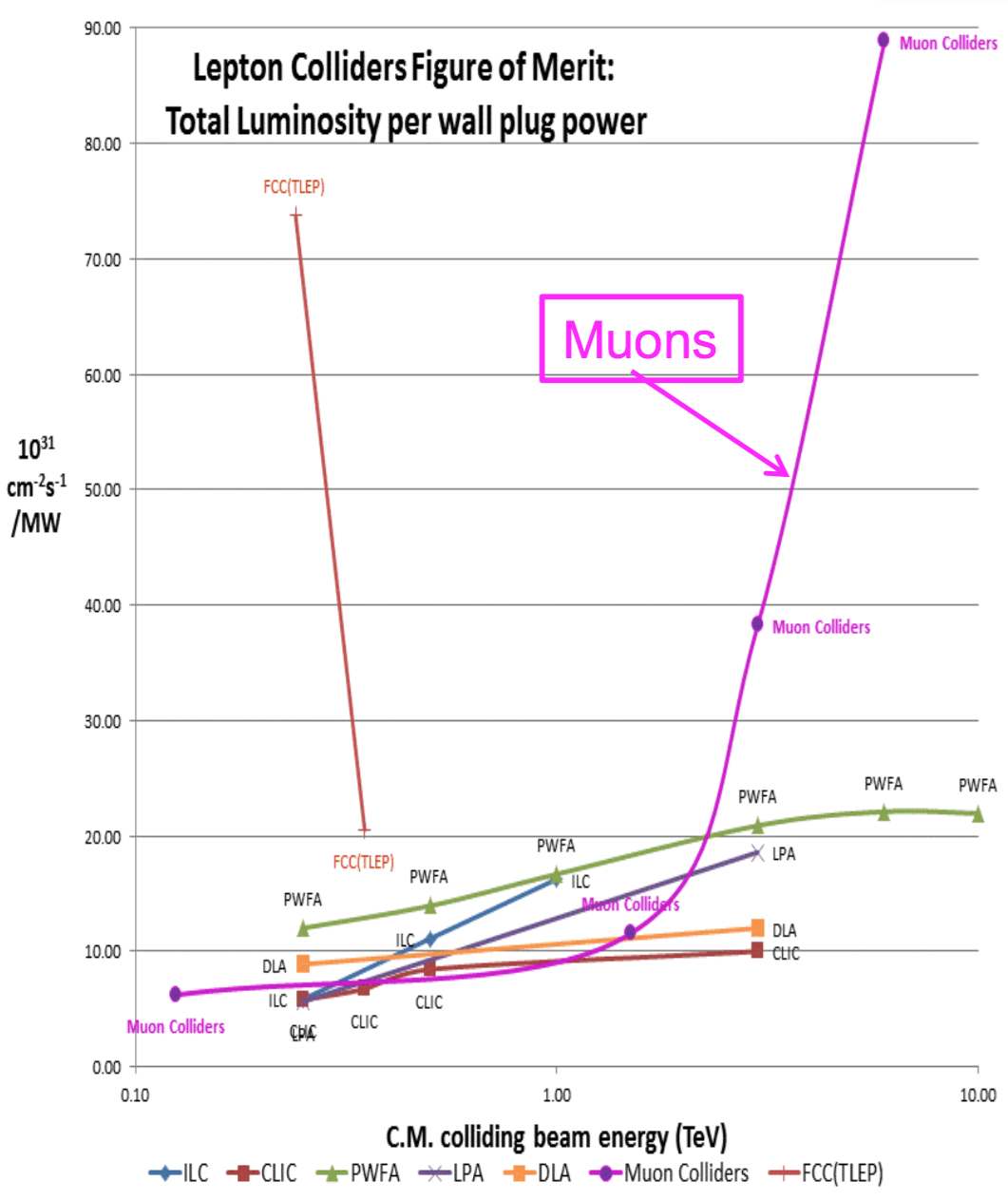}
\includegraphics[width=0.485\textwidth]{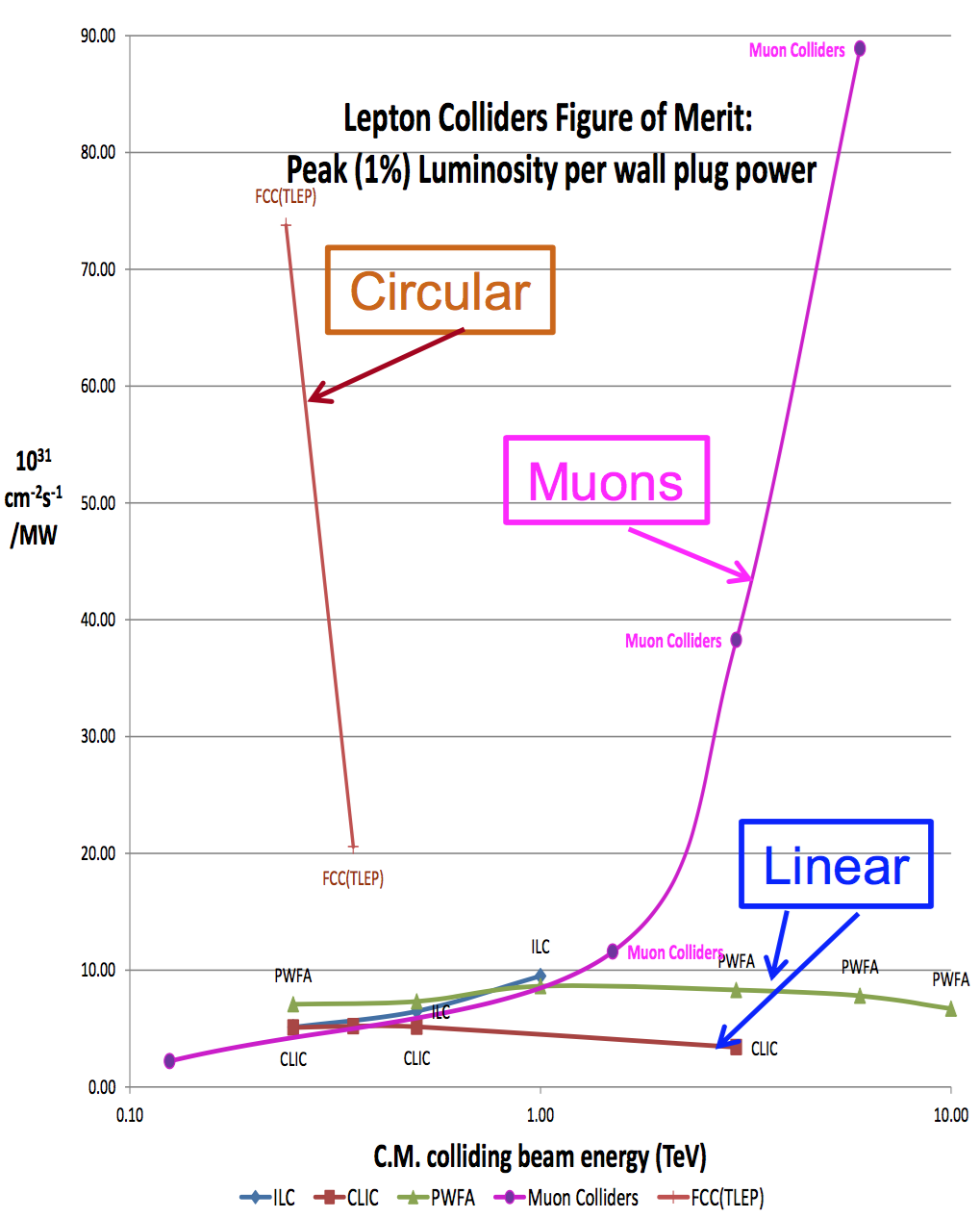}
\caption{\label{fig:lumiperMW} \small Luminosity per MW of wall plug power consumption for various lepton collider technologies~\cite{DelahayeAries2018}. Left panel: total luminosity; Right panel: luminosity for lepton-lepton collisions within $\pm 1\% $ of the nominal centre-of-mass energy. The figures-of-merit of the FCC-ee at the Z pole and the WW threshold ($1500$ and $200\times 10^{31}\,{\rm cm^{-2}s^{-1}/MW}$, respectively) are not represented in this figure.}
\end{figure}

Above 2--3\,TeV, muon circular colliders become more efficient, as shown in Fig.\ref{fig:lumiperMW}, which displays the luminosity delivered per MegaWatt (MW) for several lepton collider technologies: circular ${\rm e^+e^-}$ (FCC-ee); linear ${\rm e^+e^-}$ with superconducting RF accelerating cavities (ILC) and with a drive beam (CLIC), or with a very preliminary (and optimistic) prediction for plasma wake field acceleration (PWFA) now in an R\&D phase; and prospective circular $\mu^+\mu^-$ technology. The best figures of merit are reached by the FCC-ee at the lowest energies, up to just beyond the top-pair threshold, and by muon colliders at the highest energies, above 2.5\,TeV. Muon colliders are the only colliders for which the luminosity per MW increases significantly with the centre-of-mass energy. The PWFA technology would optimistically require half the 600 MW of CLIC at 3\,TeV for the same luminosity, but the power consumption needed to achieve the required 10 times larger luminosity at 10\,TeV, in excess of 3 GW, is much larger.  

This important drawback is worsened by the fact that, above 3\,TeV, electrons radiate their energy in the form of beamstrahlung at the interaction point and of synchrotron radiation in the focusing channels. This radiation further reduces by a large factor the effective luminosity at the nominal energy: a factor of 1.5 at 500\,GeV, a factor of 3 at 3\,TeV, possibly a factor of 10 at 10\,TeV, if only beamstrahlung is taken into account, which would need to be compensated somehow by even more wall plug power (Fig.~\ref{fig:lumiperMW}, right panel), of the order of 10\,GW or more for 10\,TeV collisions, maybe 10 times more for 30\,TeV collisions. Ultra-high energy ${\rm e^+e^-}$ colliders will therefore require much cheaper and more efficient energy production methods to be developed. 
 Muons, being heavier than electrons, are protected from these issues up to very large energies. Finally, the resulting lack of knowledge of the collision (reduced) energy and momentum, together with the important backgrounds originating from the increased $\gamma\gamma$ collisions at the interaction point, progressively reduces the usual cleanliness of electron collisions with respect to proton collisions.

{\bf In conclusion, a linear ${\rm e^+e^-}$ collider appears to be the technology of choice for lepton collisions only in the range of centre-of-mass energy extending from ~0.4 to ~2.5\,TeV, a range where new physics is currently being looked for at the LHC.} For energies above 2--3\,TeV, the circular muon collider technology seems much better suited as a solution for the far future. Circular muon collider designs are, however, much less well developed and will probably require several decades of vigorous R{\&}D to ascertain their feasibility. Finally, the usefulness of a multi-TeV muon (lepton) collider remains to be demonstrated on a case-by-case basis, depending on the physics scenario and, most importantly, on the luminosity that can be delivered by such a facility. Should the scientific interest be demonstrated, one could envisage using the 45\,GeV FCC-ee positron storage ring as an interesting first step for a multi-TeV muon collider~\cite{ZimmermannAries2018}%, after the FCC-hh
. In contrast, as muon colliders need circular tunnels, they would not be a "natural" evolution of any linear ${\rm e^+e^-}$  collider facility.

\section{Can there be a Smooth Transition between HL-LHC and FCC-ee Experiments?}
\label{sec:HLLHC}

The proposed FCC timeline integrates well with the HL-LHC schedule. The construction, installation, and commissioning of the HL-LHC detectors will take place between now and 2026. During this period, the FCC-ee study schedule foresees paper studies (detector concept developments, simulations, optimization, physics performance), and a limited amount of detector R{\&}D and beam tests for hardware concepts not already tested for the LHC, in order to be ready with Letters of Intent by the next European Strategy update. The scale of the human and financial resources needed for these detector studies is small compared to the scale of the HL-LHC collaborations, so this schedule will minimize the disturbance. 

It is only after HL-LHC operation starts in 2026 that the hard work towards the FCC-ee detector technical design will be needed, for a TDR delivery around 2031. The FCC-ee detector construction, installation, and commissioning will then take place between 2031 and the end of HL-LHC operation. Furthermore, once the construction work for HL-LHC upgrades is completed, the technical personnel, the engineers, and the detector physicists in national institutes might be transferred to activities other than HEP, or simply lost, unless there is guaranteed R{\&}D and construction work for a new generation of detectors after 2026. Once these experts are no longer fully involved with the HL-LHC, they will be most welcome to lead the work for FCC-ee detectors. Finally, FCC-ee operation will start seamlessly at the end of the HL-LHC physics programme,  ensuring the participation of high-energy physicists in HL-LHC data taking until the very end. 

High-energy physicists will want to diversify their activities as soon as the CERN Council endorses the conclusions of the European Strategy process, whatever this recommendation may be. The LEP collaborations experienced similar worries and difficulties during the LEP2 phase, when LHC was in preparation after its approval in 1994. The need for diversity is unavoidable and, in the absence of FCC-ee, there will be many possibilities to move to totally different activities, {and the know-how accumulated over decades may dissipate}.

Actually, FCC-ee is the most natural continuation of the HL-LHC programme. It would be close geographically and thematically to the HL-LHC, which makes it much easier to manage than another project not based at CERN and/or disconnected from high-energy colliders. It is therefore likely that FCC-ee activities will help keep people at CERN, safeguarding their interest in high-energy particle physics, and facilitating their involvement in the HL-LHC for a significant fraction of their time.

\section{Can Physics start at FCC-ee right after HL-LHC?}

{\bf The FCC-ee operation can start seamlessly at the end of HL-LHC.} Unlike FCC-hh, FCC-ee does not require the immediate availability of the LHC ring. The injection is done either from the SPS or directly from a dedicated Linac, and the acceleration from the injection energy to the collision energy is performed continuously with a booster inside the FCC tunnel. Hence the period of FCC-ee operation can be used to proceed with the modifications of the LHC layout and powering for future use FCC-hh injector~\cite{Milanese:2057723,Brennan:2002005}, expected to take three to five years.

\section{Will FCC-ee delay FCC-hh?}

{\bf The FCC-ee will not delay FCC-hh: instead, it will make it a realizable dream, and will maximize the significance of its physics results.} One of the great benefits of the choice of FCC-ee (with respect to any other form of lepton collider) is that it allows high-level technological effort to be concentrated on high-field magnets, while the technically simpler precursor of FCC-hh is built and operated. What appears to be additional time can be used to investigate  newer, more ambitious technologies, with the possible result of a more affordable 100\,TeV collider, or even a higher-energy collider (150\,TeV or more?). There is no doubt that the 15 years foreseen for FCC-ee operation will be put to good use in this perspective. 
 
 Moreover, the sequential implementation (first FCC-ee, then FCC-hh) maximizes the physics reach at the precision and the energy frontiers far beyond that of any linear collider project, taking advantage of the multiple complementarities and synergies of FCC-ee and FCC-hh~\cite{Mangano:2651294}. Finally, paying for FCC-hh within the timescale proposed for FCC-ee would require a substantial budgetary boost.

Other investigations for the high-energy frontier, such as the muon collider or plasma wake-field acceleration, could be pursued in parallel as accelerator R\&D projects. 

\section{How long will the Shutdown between FCC-ee and FCC-hh be?}

The schedule of the FCC integrated programme foresees 15 years of FCC-ee operation and 25 years of FCC-hh operation, interleaved with a shutdown of 10 years to dismantle the lepton collider and install the hadron collider in the FCC tunnel. This estimate for the shutdown duration results from an in-depth study based on past experience at CERN and on the planning optimization for civil engineering and infrastructure realization. It has been argued, however, that a simple extrapolation of the LEP-LHC transition to the transition from FCC-ee to FCC-hh could lead to a much longer duration~\cite{Lyn}.  

A brief account of the LEP-LHC transition period can be found in Ref.~\cite{Collier:2014zfo}. The Large Electron Positron collider was shut down on 2 November 2000, to make way for the installation of the Large Hadron Collider in the same tunnel~\cite{LEPClosureDate}, with an envisaged transition time of about four years. The LEP dismantling~\cite{LEPDismantling} started on 27 November 2000 and, after three months, the most critical two-thirds of the LEP ring had been emptied~\cite{Collier:2014zfo}. Surveying for the LHC started in November 2001 in the empty LEP tunnel~\cite{LEPSurvey}, so LEP dismantling took less than a year before work for the LHC could start. The last piece of LEP went to the surface in February 2002~\cite{LEPLastPiece}: {\bf The LEP dismantling caused no delay in the LHC installation. This experience gives no reason to believe that the FCC-ee dismantling will cause any delay to the FCC-hh installation.}

Items on the critical path to late LHC startup included the following:
\begin{enumerate}
    \item Significant infrastructure work was needed for the LHC, in particular the excavation of the new, large, caverns for ATLAS and CMS; 
    \item A financial crisis -- possibly caused by an underestimation of the LHC cost -- arose, leading to a redefinition of the cost to completion and of the commissioning schedule~\cite{LHCFinancialCrisis}, and delaying in turn the start of LHC to 2007; 
    \item The mass production of the LHC dipole cold masses was handed over to industry~\cite{LHCMassProduction} in December 2001 (i.e., after the end of LEP dismantling), and the tender was concluded in spring 2002. By December 2003, CERN had taken delivery of 154 LHC dipoles (out of a total of 1232), on which a considerable amount of testing was still necessary~\cite{LHCMagnetTesting}. 
\end{enumerate}
The installation of the cryogenic line (QRL) started in August 2003 and, after many difficulties~\cite{LHCQRL}, was complete in November 2006. The first magnet was lowered in the tunnel on 7 March 2005~\cite{LHCMagnetGreatDescent},  {\bf the full installation of the accelerator was completed in spring 2008, and the first circulating beam in the LHC was celebrated on 10 September 2008~\cite{LHCCirculatingBeam}, i.e, within three and a half years after the beginning of the magnet installation.} A major incident took place only three weeks later when a faulty electrical connection between two dipole magnets opened, triggering an electrical arc that punctured the helium vessel. The resulting high pressure helium gas wave damaged a few hundred meters of beam line, and caused the loss of approximately six tonnes of liquid helium. This incident was quickly analysed and a repair plan designed~\cite{LHCAnalysis}. This delayed the first beam in LHC as well as first collisions to the end of 2009~\cite{LHCFirstCollision}, and the real start of physics to early 2010.

\noindent The conclusion drawn from this analysis of the LEP-LHC shutdown can be summarized as follows.
\begin{itemize}
    \item As discussed in Section~\ref{sec:history}, gaining approval for the LHC was greatly facilitated by the existence of the LEP tunnel;
    \item The installation of the LHC in the LEP tunnel did not slow down the completion of LHC, but rather made it easier compared to having to excavate and complete a new infrastructure.  The LEP dismantling took less than a year. Although the LEP tunnel was initially not designed to host a 14\,TeV hadron collider, the  installation of the LHC accelerator itself, thanks to extraordinary efforts, was quite rapid, about three years. 
    {\bf A transition period of 10 years for the FCC is therefore quite a reasonable evaluation};
    \item The LHC delays during this period were largely intrinsic to the readiness of LHC itself, which was still in a preparatory phase when the LEP dismantling was over. A corollary message for the FCC-hh installation, is that {\bf the best way to ensure a short transition between two machines is to make sure that the the second one is ready to install before the first machine is shut down};
\end{itemize}
The FCC schedule is prepared in such a way as to avoid the planning- and infrastructure-related issues that made the LHC installation difficult. In particular: the tunnel diameter is much larger (5.5 m instead of 3.8 m), enabling easier installation; the large experimental caverns are to be built at the beginning of the project already for FCC-ee; the dipole magnets are being studied already today, so that mass production can start well before the initiation of FCC-hh installation; finally, FCC-ee will not be pushed to its absolute limit in the hope of finding a new particle in the last year: the transfer of scientific personnel from one FCC to the other should be much smoother. 

{\bf The planned 10-year period for the FCC-ee to FCC-hh transition takes into account the lessons learned from the LEP-LHC transition. This schedule estimate is technically solid and not aggressive.}

\section{Are there Better Ways to \texorpdfstring{100\,TeV}{100TeV} than FCC-ee?}
\label{sec:BetterWays}

This Section also includes answers to the following related questions: 
%\subsection*
\begin{itemize}
\item Should we by-pass FCC-ee and go directly for a 100 or 150\,TeV hadron collider?
\item {Should we by-pass FCC-ee and go to a high-energy upgrade of the LHC instead? }
\item {Rather than starting with FCC-ee, should we build a lower-energy hadron collider} in the FCC tunnel?
\item {Why not a low-energy linear ${\rm e^+e^-}$ collider instead?} 
\item {Should we leave FCC-ee to China?}
%\item {Why do we want the FCC in Europe?}
\end{itemize}

The keywords for answering these questions are "synergy and complementarity" and, specifically, "Why build two new infrastructures when you can do better with one?" 

The FCC study synthesized earlier grassroots initiatives dating back to 2010-2011, and was launched in 2014. It delivered four CDR volumes in December 2018, and concluded without ambiguity that the best scenario is the implementation of FCC-ee first, followed by FCC-hh. Citing the FCC-integrated-programme ESPP submission~\cite{Benedikt:2653673}:

{\sl The most efficient method for the thorough exploration of the open questions in modern physics is a staged, integrated research programme, consisting of a high-intensity lepton collider to achieve an exhaustive understanding of the Standard Model to an extent that guides the optimised design of a cost-effective and sustainable energy frontier
collider that re-uses the entire existing technical and organisational infrastructures. As was the case with LEP followed by LHC, this approach permits the control of technical and financial risks without self-imposed constraints. It cements and enlarges Europe’s leadership in particle and high-energy physics for decades to come. This program is complementary to other ongoing research activities (and leverages cross-disciplinary synergies to
expand our understanding of the universe).} 

The reasons for this efficiency are {\it (i)} the powerful physics complementarity of the two machines, leading to an unbeatable exploration potential for the integrated FCC (ee and hh) programme; {\it (ii)} the synergy of infrastructure, which leads to a considerable cost saving, and reduces the financial burden; {\it (iii)} the implementation schedule, which opens a time window of 25 years for the development of  the critical technology of high-field magnets for the hadron collider, reducing the financial and technological risks; this time window also can be well used to optimize the detector designs to match the evolution of the physics landscape; and {\it (iv)} the duration of the programme (about 70 years) and its strategic importance, make it thinkable to obtain the upfront funding for the common infrastructure.    

Proposed initially at CERN, a similar proposal exists now in China under the name of CEPC-SppC. This competing proposal makes it essential for the European Strategy Group to come to a firm recommendation for the FCC at CERN: in its absence, CERN and Europe risk losing very rapidly the leadership in high-energy physics that it presently enjoys.     

After a brief travel back in time, we discuss some financial facts and then answer the above questions, considering alternative scenarios.  

\subsection{Learning from History} 
\label{sec:history}
In the 1980's, CERN embarked on the construction of LEP, an "Electroweak Collider" with a centre-of-mass energy of 70 to 200\,GeV that would allow observation of the Z boson with unprecedented precision (including measuring the number of light neutrinos) and the W pair production process, which would have a divergent cross-section if it were not for the cancellations predicted by the Standard Model. Very soon after LEP was first discussed in 1976, it was realized that one could also place a hadron collider in the same tunnel. %In 1984, it was decided to lower LEP by 30\,cm to make space for a hadron collider (although one was far from understanding how it could be built). 
It is likely that, if the LEP tunnel had not existed, the LHC would never have happened. At the same time, with the LEP tunnel existing, the hadron collider project had an important advantage in cost (allowing construction within a constant budget) and in the approval process. It also benefited from the presence at CERN of a large community of accelerator experts and of a growing community of 1500 physicists working on the four LEP experiments. 

On the other side of the Atlantic, two projects were considered and in competition: the 40\,TeV pp collider SSC in Texas and the SLAC ${\rm e^+e^-}$ 500\,GeV linear collider project (NLC). Many explanations/interpretations have been given why the SSC was terminated while already under construction. History might have taken a different course if the SSC had been able to use a pre-existing infrastructure, for example a tunnel built as a first step for a 90-to-350\,GeV circular ${\rm e^+e^-}$ collider at an existing site such as Fermilab. Physics would probably be ahead of where it is today by at least one 20-years cycle, and perhaps two. Re-use of a common infrastructure  {\it (i)} allows a smoother funding profile, avoiding a vulnerable funding bump; {\it (ii)} prevents from having to start in a green-field site; and {\it (iii)} avoids a situation in which two communities could compete for the same resources for two different large infrastructures at the same time. In a more integrated and interdependent world, what was then an internal US issue has now become a world-wide problem of significantly larger dimensions. 

\subsection{Looking at the numbers} 
The estimated costs of the FCC programme were summarized in Section~\ref{sec:cost}.
 
It has also been estimated that the cost of the HE-LHC is 7\,GCHF, the cost of CLIC$_{380}$ is 6\,GCHF and the cost of ILC$_{250}$ is 6.5\,GCHF, so that the cost of building any of these machines in addition to that of reaching one day 100\,TeV pp collisions is greater than the 4\,GCHF marginal cost of the FCC-ee. {\bf Among all the ${\bf e^+e^-}$ or hadron colliders proposed to date, the FCC-ee, as included in an integrated FCC (ee and hh) programme, is the most cost-effective intermediate project on the path to the 100\,TeV pp collisions}. 

It is also the intermediate collider that provides the strongest physics output (Fig.~\ref{fig:FCCPhysics}). The combination of ee and hh provides a complete programme of studies of the Higgs boson, with the two machines providing almost perfect complementarity. The FCC-hh will produce more than $10^{10}$ Higgs bosons with ${\rm H}\to \mu\mu, \gamma\gamma, {\rm Z}\gamma, \upsilon\gamma, J/\psi\gamma, \phi\gamma$, or invisible decays, and with abundant samples of ttH and HH events, to complete the FCC-ee Higgs physics programme, while FCC-ee will have  provided knowledge of the Higgs total width, and an absolute and accurate determination of the $\rm HZZ $ coupling that serves as a "standard candle" to render the FCC-hh measurements model-independent. In addition, FCC-ee will provide an incomparable set of exquisite precision measurements and searches for rare Z decays and symmetry violations, while FCC-hh will have by far the most powerful reach for new physics at high energy (Fig.~\ref{fig:FCCPhysics}). 

We now turn to various alternative scenarios. 

\subsection{Should we by-pass FCC-ee and go directly for a 100 or \texorpdfstring{150\,TeV}{150TeV} Hadron Collider?}
 
Given the cost of 24\,GCHF for the FCC-hh in a standalone scenario, and  given the status of high-field magnet R{\&}D and the anticipated target cost of the magnets, it is not a realistic scenario to expect that such a machine could start operation in the 2040's. A further disadvantage is that the use of the infrastructure  would be reduced to one project, thus increasing its cost per year of foreseeable use. The opportunity to build the FCC-ee and to profit from its impressive and largely unique exploratory programme would be lost, and the physics output of the FCC-hh would be significantly diminished. 

The clear case for an ${\rm e^+e^-}$ collider would have to be satisfied elsewhere in the world. If it is a linear collider, it seems that this could not happen without a large European contribution, %at a level that is a very significant fraction of the marginal cost of the FCC-ee %{\bf find reference to Heuer suggesting Europe would contribute 2B~Euros to linear collider}, 
further limiting the ability of CERN to invest in its own infrastructure. If it were instead a circular collider in China, the likelihood that it would be followed by a hadron collider is high. In that case, CERN might miss the opportunity to build this powerful exploration machine. (The case of a circular ${\rm e^+e^-}$ collider in China is discussed more completely in Section~\ref{sec:China}.)

\subsection{Should we by-pass FCC-ee and opt for a High-energy Upgrade of the LHC instead? }

This scenario (HE-LHC) is part of the FCC study, and is the object of its 4th CDR and ESPP contribution~\cite{Zimmermann:2651305,Zimmermann:2653676}. It would consist of replacing the LHC dipoles with 16~T magnets to reach a centre-of-mass energy of 27\,TeV, doubling the mass reach for BSM searches and providing interesting prospects for Higgs physics, particularly for  ttH and HH production. The HE-LHC option could also be interesting if the HL-LHC discovered new physics at the high end of its mass reach. Doubling the LHC energy would indeed increase the production rates of the new phenomena, and allow a more detailed study. The current strong limits already obtained by the LHC, however, make even the doubling of energy a rather limited step to fully explore potential scenarios for new physics, of which a LHC discovery could just be the lowest-lying state of the spectrum. An accurate judgement of the added value of HE-LHC following a LHC discovery would therefore depend on its specific features. 

The HE-LHC has long been seen as an appealing scenario, in the absence of a realistic evaluation of its cost and inherent difficulties. The HE-LHC would be extremely constrained by  the small diameter of the LHC tunnel, 3.8 m instead of the 5.5 m diameter foreseen for the FCC tunnel. The injection energy from the SPS forces the magnets to have the same physical aperture as those of the LHC, thus increasing their cost in comparison with those that could be used in FCC-hh. An alternative possibility would be to rebuild the SPS to a higher energy machine, adding to the expense. 

The HE-LHC project cost has been estimated to 7\,GCHF, i.e. 3\,GCHF more than the marginal cost of FCC-ee, with a minimal fraction of the  infrastructure reusable for the FCC-hh. Additionally, the installation of HE-LHC would occupy the LHC infrastructure for at least 6 years after the end of HL-LHC, without collider physics at CERN. Adding an estimated 20 years of operation, followed by the adaptation of the machine to serve as an injector for FCC-hh, this would take us well into the 2060's before the 100\,TeV machine could start, which is not earlier than in the integrated FCC scenario. 

{\bf En route to 100\,TeV, the physics return that can be guaranteed by the HE-LHC is smaller than that of the FCC-ee option. Direct discoveries are clearly possible, and could modify our assessment, but otherwise the added value with respect to the HL-LHC appears inferior to the greatly complementary inputs provided to FCC-hh by the FCC-ee physics programme. The benefits of the HE-LHC path to 100\,TeV are further reduced by the additional costs,  relative to the integrated FCC plan (ee+hh).}

\subsection{Rather than starting with FCC-ee, should we build a Lower-Energy Hadron Collider in the FCC Tunnel?}

This possibility, which is not part of the FCC design study, has been raised during discussions at the European Strategy symposium in Granada, and subsequently.  It involves staging FCC-hh by using, in a first iteration of the project, magnets based on established NbTi technology~\cite{Rossi:2013xza}. 
We shall refer to this phase as LE-FCC (low-energy FCC). This approach is supported by two considerations: {\it (i)} a tested technology puts magnet construction on a fast track, bypassing the 20-year long, challenging R\&D  phase foreseen by the FCC-hh CDR; and {\it (ii)} the cost of this first stage would be lower than the full 100\,TeV collider. Studies have begun to evaluate the cost of the project, in relation to various performance scenarios for energy and luminosity. It already appears quite clear that even an LHC-like magnet design, which would allow an energy of about 48\,TeV to be reached, would lead to a cost far exceeding the cost of the FCC-ee accelerator, and beyond the budget targets that could allow operation by 2043. Important savings can be obtained by reducing the magnetic field below 6~T, leading to an energy of $37.5$\,TeV or less. Aside from the physics considerations, the assessment of the outcome of these studies should weigh the additional cost relative to the FCC-ee phase, cost that would increase the overall budget required to attain the ultimate target of 100\,TeV, putting in jeopardy the upgrade of the first stage to ultimate performance.

As we wait for the technical assessment, we focus here on some general physics considerations. The physics case of FCC-hh builds on three pillars: {\it (i)} its contribution to the Higgs, EW and top precision measurement programme; {\it (ii)} its direct discovery reach at high mass; and {\it (iii)} its potential to conclusively answer questions like whether dark matter (DM) is a thermally-produced weakly-interacting massive particle (WIMP), or whether the electroweak phase transition was of strong first order or not. {\bf For all three sets of goals, the choice of 100\,TeV as target energy plays an essential role, which, contrary to statements occasionally heard, is fully justified and required}, as briefly summarized here.

The precision Higgs measurements of FCC-hh, documented in the Physics CDR~\cite{Mangano:2651294}, benefit in several ways from high energy. On one hand, higher energy leads to larger inclusive statistics. On the other, the extended kinematic reach enables measurements with reduced systematic uncertainty, and probes Higgs interactions at large $Q^2$, where the sensitivity to deviations from the SM is enhanced and complementary to that at a low-energy $\rm e^+e^-$ Higgs factory. The precise measurement of ratios of branching ratios such as $B({\rm H\to X})/B({\rm H\to 4\ell})$ ($\rm X=\gamma\gamma,\; \mu^+\mu^-, \; Z\gamma$) will likely reach the level of per-cent precision even at $\sqrt{s}\sim 40$\,TeV, with $L\sim 10$ab$^{-1}$. But the measurement of the Higgs self-coupling will be significantly penalized, since the total rate at $\sqrt{s}=37.5$\,TeV (48\,TeV) would be smaller by a factor of 5 (3), with respect to 100\,TeV. The potentially precise determination of the ttH coupling, from the ratio of ttH/ttZ with highly boosted final states, would likewise suffer from statistics (and from the uncertain knowledge of the ttZ EW coupling, in absence of a dedicated measurement above the $\rm e^+e^- \to t\bar{t}$ threshold at FCC-ee). The resulting uncertainty on the ttH coupling would enter as dominant uncertainty in the extraction of the Higgs self-coupling from the measurement of the $\rm gg\to HH$ production rate, where the ttH coupling plays a key role in the cancellation between (triangle) self-coupling and (box) double emission diagrams. All things considered, the measurement of the Higgs self-coupling at LE-FCC would have an uncertainty at least two to three times larger than FCC-hh: 100\,TeV and 30\,ab$^{-1}$ are the minimum requirement to reach the ambitious goal of 5\%. In the context of EW precision measurements, the reduced reach in Drell-Yan mass at LE-FCC would compromise the sensitivity to the electroweak parameters $Y$ and $W$~\cite{Farina:2016rws}, bringing it below CLIC's targets.

The smaller energy would clearly imply a proportionally smaller mass reach for direct discovery at the high-mass end. Here the 100\,TeV energy is an important milestone, emerging from the indirect sensitivity to new physics promised by the $\rm e^+e^-$ colliders through their Higgs, electroweak, or flavour programmes. The energy of a future high-energy hadron collider must be scaled to allow direct discovery of new phenomena possibly revealed indirectly in $\rm e^+e^-$ collisions. Most studies presented during the European Strategy symposium~\cite{Paris} in Granada show that 100\,TeV is the energy required to achieve this goal.

Moreover, it must be stressed that not all new physics would show up via indirect precision measurements at $\rm e^+e^-$ colliders. Limits from $\rm e^+e^-$ on particles decoupled from leptons, such as squarks or gluinos, are clearly very weak. But the same can be true of weakly interacting particles. For these particles (such as supersymmetric charginos, neutralinos or sleptons), LEP itself did not set very strong limits, beyond those set by the search for direct production.  For this reason, the parameters of a future hadron collider must also be tuned to push further the search for new particles just above the thresholds of future $\rm e^+e^-$ colliders. Even if the masses here are not large, energy remains crucial, to achieve production rates large enough to guarantee discovery, in view of the otherwise small cross sections. It has been shown in the FCC Physics CDR~\cite{Mangano:2651294} that Wino and Higgsino DM candidates can be discovered or ruled out up to the masses of $\sim 4$ and $\sim 1.5$\,TeV, just above the general upper mass limits, $\sim 3$ and $\sim 1$TeV respectively, dictated by cosmology. A factor of two reduction in centre-of-mass energy would miss the target of conclusively discovering, or ruling out, such WIMP candidates. In a similar way, studies of the direct signals for a strong first order electroweak phase transition, documented in the FCC Physics CDR, show that 100\,TeV (and 30ab$^{-1}$) are necessary to cover the full range of scenarios.

In conclusion:  100\,TeV does represent an important energy threshold to guarantee an important set of deliverables, from the measurement of the Higgs self-coupling at the $\sim 5$\% level, to the exploration of WIMP DM, of the nature of the electroweak phase transition, to the search for particles responsible for deviations observed by precision $\rm e^+e^-$ measurements. As such, 100\,TeV or more should remain as an ultimate target of the FCC programme. Lower energies would certainly allow progress with respect to the LHC, but, contrary to FCC-ee, would not add complementary information to the ultimate output of FCC-hh, and might weaken the physics case for FCC-hh by pre-empting some of the FCC-hh measurements. The cost of transiting through a lower-energy option, furthermore, would very likely add to the final cost of achieving 100\,TeV, possibly jeopardizing this eventual upgrade altogether. {\bf Altogether, the route to 100\,TeV via FCC-ee seems a lot more promising.}  
%}

\subsection{Why not a Low-Energy Linear \texorpdfstring{${\bf e^+e^-}$}{ee} Collider instead?}  
It should be emphasized at this point that the physics capabilities of linear ${\rm e^+e^-}$ colliders on the one hand,  and of circular electroweak, flavour, and Higgs factories (FCC-ee or CEPC) on the other, are not identical; a significant amount of complementarity exists between them.  Although they overlap in the 240-380\,GeV centre-of-mass region, there are considerable differences otherwise. The TeraZ part of the programme is unique to circular colliders and, in the present physics landscape, provides great opportunities for discoveries pointing to the existence and nature of the BSM physics still needed for the unexplained features of the Universe. The linear colliders are unique in their ability to reach energies of up to 2--3\,TeV centre-of-mass with lepton collisions, though this physics scope overlaps significantly with that of the hadron colliders, including already the LHC. While their specific physics case might be greatly enhanced with the discovery of a new particle or phenomenon in their mass range, linear colliders are largely hunting on the same grounds as hadron colliders in many searches or measurements. 

The two above scenarios, HE-LHC and LE-FCC, in which CERN would specialize indefinitely in hadron colliders, could find favour among partisans of a non-synergistic linear ${\rm e^+e^-}$ collider built elsewhere in the world, presumably Japan. It is worth emphasizing, however, that the costs of these scenarios would be considerably higher for the global particle physics community than the integrated FCC programme%, subject to offsets due to host-state contributions to ILC and/or FCC
. This is true  even in the most minimal scenario of a 250\,GeV collider, and would become dramatic in the case of upgrades in energy. 

From the physics point of view, while a 250\,GeV linear collider would fulfil some of the important tasks of FCC-ee (such as the measurement of the Higgs boson width and its coupling to the Z), it would do so with somewhat lesser precision and over a considerably longer exposure (Section~\ref{sec:LowEnergyHiggsFactory}). Furthermore, it would reduce the overall physics scope available in the world because it would offer much-reduced capabilities for Z running, thus losing out on the TeraZ programme that provides a substantial part of the FCC-ee discovery potential.  As explained before, the combination of FCC (ee and hh) is by far the most powerful and complete way to cover Higgs, electroweak, and flavour physics and the wider scope of exploration.

For these reasons, {\bf if ILC were to proceed, the impact of Europe on high-energy physics and the physics complementarity of the different phases of the  FCC would still argue strongly in favour of pursuing the full FCC route (ee and hh).} 

\subsection{Should we leave FCC-ee to China?}
\label{sec:China}

As said earlier, the integrated FCC programme is reflected in a similar proposal  in China under the name of CEPC-SppC, with a rapid technical schedule that foresees physics starting in 2029, under the assumption of an ideal funding scenario. 

The consequences for particle physics in Europe of the realization of this project would be considerable. All the technical and scientific qualities of FCC would be shared by this essentially identical project, which has thus the broadest, most encompassing physics discovery reach, with respect to which it would be extremely difficult to maintain leadership. As with ILC, however, the information we have about the CEPC planning does not support the expectation that this project will be formally approved for construction during the time-frame of the European Strategy update process, and its chances of eventual approval cannot obviously be guaranteed today. 

In this situation, the European Particle Physics community has a chance to decide on its own future independently, and to remain at the forefront of high-energy physics, without being pre-empted by putative scenarios of future plans in other regions. 
More general motivations are elaborated in the following Section. This decision should be based on the scientific judgement of what is best for the future of HEP from the European perspective. The concrete exploration of political and financial scenarios, leading to a formal approval of the next collider at CERN in a time frame that will allow operation to start upon the completion of the HL-LHC, requires the European Strategy for Particle Physics (ESPP) Update to emerge with a clear recommendation in 2020. 

A firm resolution on a European project at CERN beyond the LHC would represent the first concrete commitment, on a worldwide scale, to engage in a future collider project. This commitment would constitute a solid lead for other regions deciding how best to invest their resources, with an opportunity for Europe to contribute to their ambitions through reciprocal collaboration, while maintaining the priority of Europe's targets.

In conclusion, {\bf the appearance of an ambitious alternative makes it essential for the ESPP to express in 2020 a firm recommendation for the FCC at CERN. In its absence, CERN's future becomes a hostage to fortune and Europe may have to fall back on a less powerful instrument, with the risk of losing very rapidly the leadership in high energy physics that it presently enjoys.  On the contrary, a clear European position would constitute a focus for a worldwide collaboration. }

\section{Why do we want FCC in Europe?}

The integrated FCC is the most visionary proposal that fulfils the recommendation of the European Strategy in 2013:
{\em ``To stay at the forefront of particle physics, Europe needs to be in a position to propose an ambitious post-LHC accelerator project at CERN by the time of the next Strategy update"}. 
%The choice of the e+e- circular collider as a first step is dictated by the physics landscape. The size of 80-100\,km is optimal for studying the heavy particles of the Standard Model with an e+e- collider, and guarantees a big jump in energy reach for a hadron collider. It is the most powerful long term vision that has been designed today.
Alternative facilities that are proposed as providing a similar programme of Higgs studies, are less precise; not much cheaper; and considerably less broad in physics perspectives. As seen in Section~\ref{sec:BetterWays}, the other routes to reach 100\,TeV pp collisions are less precise, less complete, and more expensive.

Over the past 65 years, step by step and exploiting synergies between successive accelerators, Europe has developed a laboratory, CERN, that is now leading the field. With its demonstrated extraordinary competence, its international membership, its built-in cooperation among countries sharing common ideals of freedom and democracy, and the existing infrastructures (accelerator and injector complex, cryogenics, mechanics, electronics, workshops, and its many competences), CERN is the best place for a challenging enterprise such as FCC.

The FCC CDR makes a compelling list of the benefits for all CERN member states, and more generally for all participating countries, of hosting the FCC project at CERN. Such benefits encompass technological and industrial applications in fields that range from information technology to fast electronics, particle accelerator and detector technologies and know-how, which in turn are put into practice in communications, medicine, health, and many other sciences or day-to-day use. The combined LEP/LHC/HL-LHC cost-benefit analysis~\cite{Bastianin:2319300,Bastianin:2635876,Florio:2015dna,Florio:2016uma} revealed that a long-term programme consisting of a technology-ready lepton collider (FCC-ee), followed by a highest energy hadron collider (FCC-hh) is most likely to generate the highest possible socio-economic impact. It is therefore, {\bf not only for the physics, but also from a socio-economic point of view, an unbeatable scenario~\cite{Benedikt:2653673}}.

To lead in the race to knowledge and discovery is a tremendous motor for ingenuity and excellence, that permeates the whole of society,
particularly young people.  CERN has developed extraordinary experience, in a collaborative context, in running these large scientific projects based on world-wide collaborations of hundreds of institutes. CERN is the home of European particle physicists, and it is also the radiating centre of an intense worldwide collaboration.  

The physics we do today, hand in hand with cosmology, astrophysics and many other fields,
addresses questions -- {\it How was the Universe born and how does it work?} -- that have fascinated humanity and raised enthusiasm for a very long time. The existence in Europe of the fulcrum of a large community of scientists from all continents, genders, cultures, and religions, working together to address to these fundamental questions, with explanations based on facts, in a language that is universal, is a tremendous hope for education, world harmony, and peace. Progress in knowledge has no price.

As Europeans, we can be proud that Europe hosts such a place and we should strive to keep it at the forefront of the worldwide effort.

\bibliographystyle{jhep}
\bibliography{sample}

\end{document}